\documentclass[sn-mathphys, icol]{sn-jnl}
\usepackage[utf8]{inputenc}
\usepackage{xcolor}
\usepackage{amsmath}
\usepackage{amssymb}
\usepackage{mathtools}
\usepackage{amsthm}
\usepackage{float}
\usepackage{rotating}
\usepackage{amsmath}
\usepackage{hyperref}
\usepackage{tabularx}
\usepackage{graphicx}
\usepackage{caption}
\usepackage{subcaption}
\usepackage{booktabs} 
\usepackage{algorithm,algpseudocode}
\usepackage{array}
\usepackage{enumitem}
\usepackage{lmodern}
\usepackage{cancel}
\usepackage{arydshln}

\usepackage[square,numbers,sort&compress]{natbib}
\theoremstyle{plain}
\newtheorem{theorem}{Theorem}[section]
\newtheorem{proposition}[theorem]{Proposition}

\newtheorem{corollary}[theorem]{Corollary}
\theoremstyle{definition}

\theoremstyle{remark}

\newcommand{\ewVar}{\boldsymbol{v}}
\newcommand{\TotVar}{\boldsymbol{v}}
\DeclareMathOperator{\Tr}{Tr}

\newcommand{\x}{\boldsymbol{x}}
\renewcommand{\v}{\boldsymbol{v}}

\usepackage{todonotes}
\usepackage{cleveref}

\usepackage{lineno}
\newcolumntype{Y}{>{\centering\arraybackslash}X}

\title[]{Controlling Ensemble Variance in Diffusion Models: An Application for Reanalyses Downscaling}

\author*[12]{\fnm{Fabio} \sur{Merizzi}}\email{fabio.merizzi@unibo.it}
\author[1]{\fnm{Davide} \sur{Evangelista}}\email{davide.evangelista5@unibo.it}
\author[2]{\fnm{Harilaos} \sur{Loukos}}

\affil[1]{\orgdiv{Department of Informatics: Science and Engineering (DISI)}, \orgname{University of Bologna}, \orgaddress{\street{Mura Anteo Zamboni 7}, \city{Bologna}, \postcode{40126},
\country{Italy}
}}

\affil[2]{\orgdiv{The Climate Data Factory (TCDF)}, \city{Paris}, \country{France}
}

\abstract{In recent years, diffusion models have emerged as powerful tools for generating ensemble members in meteorology. In this work, we demonstrate how a Denoising Diffusion Implicit Model (DDIM) can effectively control ensemble variance by varying the number of diffusion steps. Introducing a theoretical framework, we relate diffusion steps to the variance expressed by the reverse diffusion process. Focusing on reanalysis downscaling, we propose an ensemble diffusion model for the full ERA5-to-CERRA domain, generating variance-calibrated ensemble members for wind speed at full spatial and temporal resolution. Our method aligns global mean variance with a reference ensemble dataset and ensures spatial variance is distributed in accordance with observed meteorological variability. Additionally, we address the lack of ensemble information in the CARRA dataset, showcasing the utility of our approach for efficient, high-resolution ensemble generation.}
\begin{document}

\maketitle

\noindent \textbf{keywords:} Diffusion Models,  Ensemble Generation, Reanalysis Downscaling, Calibrating Variance

% \markboth{}{}
% \newpage
\section{Introduction}\label{sec:Introduction}
Neural approaches to downscaling meteorological reanalyses have proven effective, enhancing spatial and temporal resolutions while reducing computational costs compared to traditional numerical models.

The task is typically framed as an image-to-image prediction, transforming a low-resolution input into a high-resolution output. Widely adopted prediction-based models, such as U-Net \cite{ronneberger2015} and ViT \cite{liu2021swin}, operate deterministically and predict the most likely outcome for a given input. While this can produce high-quality predictions, these models lack the ability to quantify uncertainty or capture extreme events.
To address this limitation, conditioned generative approaches, such as diffusion models \cite{ho2020denoising}, have gained traction in the scientific community. Unlike deterministic models, diffusion models sample outcomes from a learned distribution, producing diverse ensemble members even when conditioned on the same low-resolution input. These ensembles provide valuable insights into uncertainty estimation, the underlying physical system, and extreme events.

In this paper, we apply a Denoising Diffusion Implicit Model (DDIM) \cite{song2022denoisingdiffusionimplicitmodels} to downscale meteorological reanalyses, focusing on wind speed. Specifically, we perform the transformation from a global reanalysis (ERA5 \cite{Hersbach_ERA5_2018}) to a regional reanalysis (CERRA \cite{ridal2024cerra}) over the European domain. Building on prior work \cite{Merizzi2024}, we develop a diffusion model capable of operating across the full CERRA domain and we explore architectural modifications to scale diffusion models to large domains.
A critical aspect of generative diffusion is the statistical properties of the generated distribution, particularly the variance. We aim to mathematically analyze how variance relates to the model's architecture and ensure its spatial distribution aligns with physical expectations.
Our analysis begins with a mathematical exploration of the reverse diffusion process, revealing that the variance of ensemble members relates to the number of diffusion steps. Although previous works have already reported a reduction in variability when using fewer diffusion steps \cite{song2022denoisingdiffusionimplicitmodels, nichol2021improved} the best of our knowledge, no study has systematically investigated the relationship between ensemble variance and the number of diffusion steps. We experimentally validate this relationship in the ERA5-to-CERRA task, leveraging the CERRA-EDA \cite{ridal2024cerra} ensemble dataset for comparison. Our results demonstrate that adjusting the number of diffusion steps provides sufficient control to match the variance expressed by CERRA-EDA. This establishes our method as a practical tool for tuning ensemble diffusion to align with known uncertainty.
Finally, studying the spatial distribution of the variance, we provide a methodology to obtain the optimal number of diffusion steps in specific applications, and we showcase our approach by applying it to a dataset of the Artic region lacking ensemble information, CARRA \cite{Copernicus_Arctic_Reanalysis}.

\subsection{Background}\label{subsec:reanalysis_downscaling_DL}
Reanalysis datasets are a cornerstone of modern climate and weather research, providing gridded, high-resolution reconstructions of past atmospheric conditions. Reanalysis are widely used by researchers, institutions, and organizations to support studies ranging from climate variability to extreme event prediction \cite{rs12060980,MOREIRA2019335,zandler2020reanalysis}. High-resolution reanalyses are often created using downscaling techniques, which refine coarse global climate data into finer spatial and temporal scales. Traditional dynamical downscaling is computationally expensive, requiring significantly more resources than coarser global climate models and leading to data gaps and incomplete assessments of model uncertainty and regional climate variability \cite{Pierce2009,Goldenson2023}.

In recent years, deep learning has emerged as an effective alternative for downscaling, offering powerful and cost-effective methods for producing high-resolution data \cite{Sun2024,perez2024}. A variety of deep learning models for downscaling have been proposed \cite{vandal2017deepsd,cheng2022,rs14030769,Ji2020Downscaling,Harris2022,passarella2022reconstructing,quesada2022repeatable,ZHANG2021119321,nguyen2023climatelearn,Harder2022Generating,dujardin2022wind,9932682,price2023gencast}, showcasing their ability to capture complex spatial patterns and generate high-quality outputs across a range of meteorological variables. 

Among these methods, diffusion models stand out for their ability to capture the uncertainty of the downscaling process, enabling not only the generation of high-resolution data but also produce probabilistic ensembles by sampling from a latent space. This capability makes them highly effective for modeling uncertainty and variability, which are crucial in meteorological applications \cite{mardani2024,lopezgomez2024,ling2024diffusion,Li2024,wan2023debias, asperti2023precipitation, leinonen2023latent}.

On the other hand, despite their early success in generative tasks, Generative Adversarial Networks (GANs) have proven to be highly unstable and difficult to generalize \cite{saad2024survey, wiatrak2019stabilizing}. Their adversarial training often leads to mode collapse, where the generator produces limited variability, and convergence is not theoretically guaranteed \cite{cobbinah2025diversity}. Furthermore, GAN-based approaches tend to be problem-specific, requiring extensive architectural and hyperparameter tuning to maintain stability \cite{saxena2021generative}, and numerous variants have been proposed merely to address training issues. As a result, GANs are being increasingly phased out in climate and meteorological downscaling tasks \cite{rampal2024enhancing, dhariwal2021diffusion} and are being superseded by diffusion models, which offer stable likelihood-based training, controllable stochasticity, and consistent convergence behavior.

Downscaled ensemble datasets are essential for quantifying uncertainties \cite{Wootten2020The,lopez2024dynamical,Lockwood2024}, detecting extreme events \cite{trenberth2015attribution}, and performing ensemble-based statistical analyses critical for assessing the likelihood of compound extreme events \cite{Kornhuber2023,Anderson2019}.
In recent years an increasing amount of publications focused on the utilization of generative deep learning architectures in the downscaling task, utilizing the inherently probabilistic nature of the models to generate ensemble members.

{Recent studies have explored the use of generative models for meteorological downscaling and ensemble generation.  
In \cite{Harris2022}, generative adversarial networks (GANs) were extended to downscale precipitation forecasts, producing spatially coherent maps while correcting forecast errors.  
Similarly, \cite{rs14030769} applied a dual-learning GAN framework to satellite sea surface wind (SSW) data, combining high-resolution reconstruction with degradation kernel estimation.  
While these GAN-based approaches demonstrated promising results, they remained limited in their ability to model uncertainty.  }

{More recently, diffusion-based generative models have gained attention for probabilistic weather forecasting and downscaling.  
\cite{Li2024} employed diffusion models to generate ensemble forecasts that replicate the statistical properties and predictive skill of operational systems, while improving bias correction and reliability.  
In \cite{Lockwood2024}, a conditional diffusion model was proposed for tropical cyclone wind-field downscaling, achieving realistic high-resolution structures and ensemble variability.  
Finally, \cite{lopez2024dynamical} combined dynamical and diffusion-based downscaling to enhance regional climate projections, improving uncertainty estimates and multivariate correlations.}

{Together, these works highlight the growing relevance of diffusion models for meteorological ensemble generation.  
However, none of them explicitly investigate how the variance of generated ensembles can be controlled or calibrated—a gap that the present work addresses through a theoretical and empirical analysis of the relationship between variance and the number of diffusion steps.}

\subsection{Diffusion models basics}\label{subsec:introduction_DDIM}
Diffusion models, and in particular Denoising Diffusion Implicit Models (DDIM) \cite{song2022denoisingdiffusionimplicitmodels}, are state-of-the-art generative models able to generate highly realistic data from noise. In particular, given a sequence $\{ \alpha_t \}_{t = 1}^T \subseteq (0, 1)$ such that $\alpha_0 \approx 0$ and $\alpha_T \approx 1$, called \emph{noise schedule}, a DDIM defines an iterative procedure to turn random noise $\x_0 \sim \mathcal{N}(\boldsymbol{0}, \boldsymbol{I})$ into a new sample $\x_T \sim p_{gt}(\x_T)$, where $p_{gt}(\x_T)$ represents the probability distribution from which the training data is sampled\footnote{Note that here we use a slightly different notation compared to e.g. \cite{song2022denoisingdiffusionimplicitmodels}, as we consider the time running \emph{reversely}. We made this choice as this notation simplifies the mathematical derivation of our formula in \cref{subsec:signal_noise_rates}.}. This procedure is obtained through a simple yet effective idea: let $\{ \x_0, \x_1, \dots, \x_{T-1}, \x_T\}$ be a sequence such that $\x_0 \sim \mathcal{N}(\boldsymbol{0}, \boldsymbol{I})$, $\x_T \sim p_{gt}(\x_T)$, and $\x_t \sim p_t(\x_t | \x_T)$ for any $t = T-1, \dots, 0$. If $p_t(\x_t | \x_T) = \mathcal{N}(\sqrt{\alpha_t} \x_T, (1 - \alpha_t)\boldsymbol{I} )$, then:
\begin{align}\label{eq:forward_process}
    \x_t = \sqrt{\alpha_t} \x_T + \sqrt{1 - \alpha_t} \boldsymbol{\epsilon}_t, \quad \boldsymbol{\epsilon}_t \sim \mathcal{N}(\boldsymbol{0}, \boldsymbol{I}).
\end{align}
Note that the previous equation implies that $\x_t$ is obtained by interpolating real data $\x_T$ with random noise $\boldsymbol{\epsilon}_t$, with interpolation coefficients $\sqrt{\alpha_t}$ and $\sqrt{1 - \alpha_t}$. Therefore, they are usually referred to as \emph{signal rate} and \emph{noise rate}, respectively. Please refer to \cref{subsec:signal_noise_rates} for a detailed discussion on how these terms are selected. \\

Clearly, Equation \eqref{eq:forward_process} represents the process that slowly corrupts a true image $\x_T$ into pure noise $\x_0$. To revert this process and generate data from noise, a DDIM considers a neural network $\boldsymbol{\epsilon}_\Theta^{t}(\x_{t})$ trained to extract the noise component $\boldsymbol{\epsilon}_t$ from $\x_t$, i.e.
\begin{align}\label{eq:DDIM_approximates_noise}
    \boldsymbol{\epsilon}_\Theta^t(\x_t) \approx \boldsymbol{\epsilon}_t, \quad \forall t = T-1, \dots, 0.
\end{align}
Given that, one can simply invert the process in \Cref{eq:forward_process} by sampling $\x_0 \sim \mathcal{N}(\boldsymbol{0}, \boldsymbol{I})$ and then iteratively updating:
\begin{align}\label{eq:generative_process}
    \x_t = \sqrt{\frac{\alpha_t}{\alpha_{t-1}}}\x_{t-1} + \left( \sqrt{1 - \alpha_t} - \sqrt{\frac{\alpha_t(1 - \alpha_{t-1})}{\alpha_{t-1}}}\right) \boldsymbol{\epsilon}_\Theta^{t-1}(\x_{t-1}).
\end{align}
After $T$ steps, this process leads to a new sample $\x_T \sim p_{gt}(\x_T)$. For a detailed introduction to DDIM, please refer to \cite{song2022denoisingdiffusionimplicitmodels,asperti2023image,karras2022elucidating}.

A property of DDIM that is crucial for our work is the possibility of \emph{skipping} a few steps in the generative process. This is achieved by setting a \emph{step-size} $\Delta t \in \mathbb{N}$ and modifying
\Cref{eq:generative_process} as:
\begin{align}\label{eq:generative_process_timestep}
    \x_t = \sqrt{\frac{\alpha_t}{\alpha_{t-\Delta t}}}\x_{t-\Delta t} + \left( \sqrt{1 - \alpha_t} - \sqrt{\frac{\alpha_t(1 - \alpha_{t-\Delta t})}{\alpha_{t-\Delta t}}}\right) \boldsymbol{\epsilon}_\Theta^{t-\Delta t}(\x_{t-\Delta t}).
\end{align}

Note that the choice $\Delta t = 1$ recovers the original generative process, while $\Delta t > 1$ allows for a faster generation, possibly at the expense of quality. This modification can be relevant when the neural network model $\boldsymbol{\epsilon}_\Theta^{t}(\x_{t})$ is particularly large, or when the dimensionality of the data $\x_T$ is big. Indeed, when this happens, computing $\boldsymbol{\epsilon}_\Theta^{t-\Delta t}(\x_{t-\Delta t})$ could be time-prohibitive, and reducing the amount of time this operation has to be done to reach the end of the process is crucial for its applicability. Since the number of iterations of \Cref{eq:generative_process_timestep} equals to $N = \frac{T}{\Delta t}$, increasing $\Delta t$ leads to a faster generation process. 

In this work we argue that the importance of the selection of the step-size $\Delta t$ is not limited to the time-efficiency of the generative process, but it can also be used to control the variance of the sampled data: a key aspect when these models are employed for generating ensembles.

\subsection{Notations}
Throughout this article, we will make use of some notations, which we report here for completeness. In particular, we use lower-case non-bold latin characters to indicate scalar values, lower-case bold latin characters to indicate vectors, and upper-case bold latin characters to indicate matrices. Moreover, we will denote as $\boldsymbol{0}$, $\boldsymbol{1}$ the vector of all zeros and all ones, respectively, and as $\boldsymbol{I}$ the identity matrix. If $\x$ and $\boldsymbol{y}$ are two vectors with the same dimension, then $\x \geq \boldsymbol{y}$ means that $x_i \geq y_i$ for every $i$. The data is therefore assumed to be a vector of dimension $n$, which corresponds to the flattened version of the $(S \times M \times h \times w)$-dimensional array if it is indicated with a lowercase bold letter, while it is treated as a tensor if indicated with an uppercase bold letter. For example, if $\boldsymbol{X} \in \mathbb{R}^{S \times M \times h \times w}$ represents the usual tensor-shaped dataset with $S$ datapoints, $M$ channels, and spatial dimension $h \times w$, then $\boldsymbol{x} \in \mathbb{R}^n$ represents its flattened version, with $n = S \cdot M \cdot h \cdot w$.

\subsection{Structure of the article}
The remainder of this paper is organized as follows. In \Cref{sec:Data} we introduce the datasets employed for the experiments, namely CERRA, CERRA-EDA and CARRA, representing a high-resolution world map with positional information of the wind. In \Cref{sec:Methodology} we provide the theoretical background for the proposed idea, and show a possible interpretation of the observed phenomena. Next we introduce the methodology applied in the successive experiments, described in \Cref{sec:Experiments}. Finally, in \Cref{sec:Conclusion} we conclude the paper by summarizing our results. 

\section{Data}\label{sec:Data}
\begin{figure}[htbp]
    \centering
    % First subfigure
    \begin{subfigure}[b]{0.50\textwidth}
        \centering
        \includegraphics[width=\textwidth]{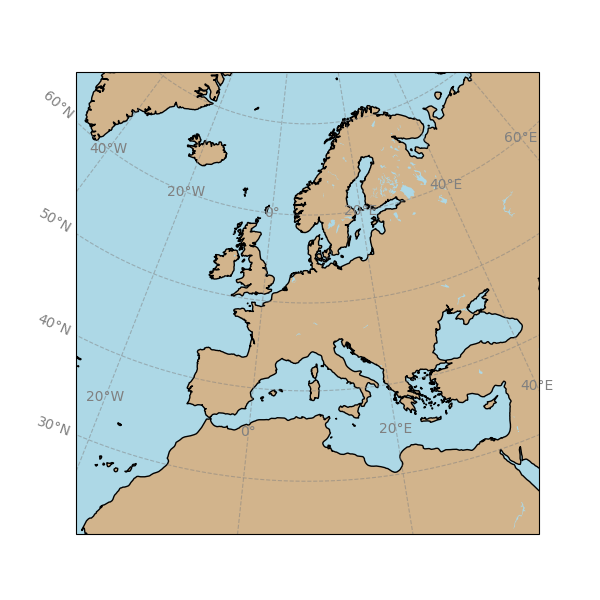}
        \caption{Domain of CERRA and CERRA-EDA}
        %\label{fig:domain-CERRA}
    \end{subfigure}
    \hfill
    % Second subfigure
    \begin{subfigure}[b]{0.39\textwidth}
        \centering
        \includegraphics[width=\textwidth]{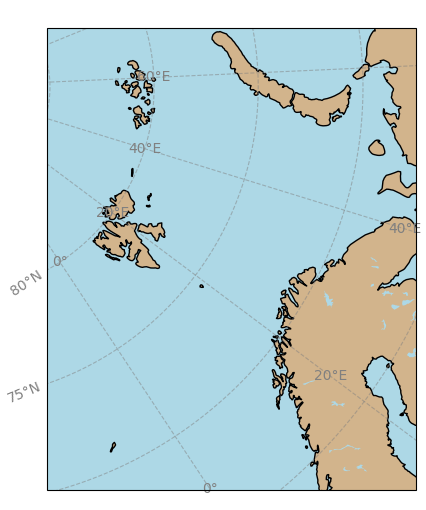}
        \caption{Domain of CARRA-EAST}
        %\label{fig:domain-CARRA}
    \end{subfigure}
    \caption{Comparison of CERRA and CARRA-EAST domains.}
    \label{fig:both_domains}
\end{figure}

In this section we describe the primary datasets used in this study. Our task involves using the global reanalyses ERA5 \cite{hersbach2020era5} as the low-resolution input to condition our model in predicting a high-resolution downscaled counterpart, trained on the regional reanalyses CERRA \cite{ridal2024cerra}. The resulting ensemble members are validated against CERRA-EDA, an ensemble version of CERRA at half the temporal and spatial resolution. We will also apply our methods to CARRA, a dataset built on the same core model as CERRA and that covers the Arctic region, but lack ensemble members. The respective domains of CERRA and CARRA are reported in Figure \ref{fig:both_domains}. This section begins with a brief introduction to ERA5, followed by a focused discussion on CERRA, CARRA, and CERRA-EDA.

\subsection{ERA5}
ERA5 \cite{Hersbach_ERA5_2018}, the fifth generation ECMWF reanalysis, is a globally recognized dataset extensively utilized in climate and atmospheric research. Spanning from 1940 to the present, it provides hourly estimates of atmospheric, land, and oceanic variables at a 0.25° horizontal resolution (approximately 30 km) and 137 vertical levels. Based on the ECMWF Integrated Forecasting System, ERA5 incorporates advanced numerical weather prediction models and satellite data, ensuring high reliability and precision \cite{hersbach2020era5}. Its widespread use spans diverse applications, including precipitation trends, temperature analysis, wind studies, and extreme event research, making it an indispensable resource in climate science.

\subsection{CERRA}\label{subsec:CERRA}
The Copernicus Regional Reanalysis for Europe (CERRA \cite{ridal2024cerra}) is a sophisticated high-resolution Regional ReAnalysis (RRA) dataset specifically designed for the European region. It is a product of the European Copernicus Climate Change Service (C3S), executed through a contract with the Swedish Meteorological and Hydrological Institute (SMHI), in collaboration with subcontractors Meteo-France and the Norwegian Meteorological Institute. CERRA operates at a high 5.5 km horizontal resolution, offering enhanced spatial detail for meteorological variables across Europe.

% \begin{figure}[htbp]
%     \centering
%     \includegraphics[width=0.45\textwidth]{domain.png}
%     \caption{Domain of CERRA and CERRA-EDA}
%     \label{fig:domain-CERRA}
% \end{figure}

To achieve this fine resolution, CERRA relies on the global ERA5 reanalysis for initial and boundary conditions. In addition to ERA5, CERRA incorporates an extensive array of high-resolution observational data, covering conventional and satellite sources, along with physiographic datasets that represent surface characteristics. One innovative aspect of CERRA is its integration with the CERRA-EDA (Ensemble of Data Assimilation), which is a 10-member ensemble system that provides essential background error statistics. These statistics, recalculated every second day, enable CERRA to dynamically adjust initial conditions, significantly enhancing the accuracy of data assimilation and model outputs.

% CERRA’s outputs encompass both forecasts and reanalyses. Forecasting involves analyzing the current atmospheric, terrestrial, and marine surface states to predict future conditions. These forecasts are generated using numerical models, initiated from the assimilated atmospheric state (the “first guess”), and refined through data assimilation—an iterative process that aligns the model more closely with recent observations. By using both high-frequency reanalyses and forecasts, CERRA ensures an accurate historical record of weather conditions, continuously aligned with observational data.

CERRA provides a comprehensive range of meteorological variables, including temperature, humidity, wind speed and direction, precipitation, and cloud cover, all presented on a consistent $1069 \times 1069$ grid. Unlike ERA5’s hourly structure, CERRA’s temporal resolution combines reanalysis and forecast data, featuring eight reanalysis time points daily at three-hour intervals starting from 00 UTC. Each reanalysis cycle is followed by a six-hour forecast period, with additional 30-hour forecasts generated at 00 UTC and 12 UTC. For this study, we focus exclusively on the reanalysis data. CERRA employs a Lambert Conformal Conic (LCC) projection to minimize distortion at high latitudes, ensuring greater accuracy in polar regions.

% This schedule enables users to choose between reanalysis data—which integrates corrective steps using observations—and forecast data, providing flexibility and continuity across the day. Consequently, while CERRA offers data at each hour, the quality of the 8 reanalysis time points is generally higher due to their direct alignment with observational corrections.

\subsection{CARRA}\label{subsec:CARRA}
The C3S Arctic Regional Reanalysis (CARRA) dataset \cite{Copernicus_Arctic_Reanalysis} addresses the unique environmental and climatic conditions of the Arctic region, which are insufficiently captured by global reanalyses such as ERA5. While ERA5 provides a 31 km global resolution, CARRA offers 2.5 km horizontal resolution, capturing finer details essential for understanding the Arctic’s rapidly changing climate. Observational records and climate projections show that warming in the Arctic occurs at over twice the global average rate, leading to increased environmental and economic activities and driving a need for detailed data to support climate adaptation and management in the region.

CARRA leverages the same HARMONIE-AROME weather prediction model as CERRA, enhanced for reanalysis applications at the ECMWF high-performance computing facility. It features a three-hourly analysis update with a comprehensive three-dimensional variational data assimilation scheme, incorporating diverse data sources. This includes an extensive array of local observations from Nordic countries and Greenland, advanced satellite-derived glacier albedo data, improved sea ice and snow initialization, and high-resolution physiography and orography data, specifically adjusted for the Arctic’s unique topography. 

The dataset spans two key domains:

\begin{itemize}
    \item CARRA-WEST, covering Greenland, the Labrador Sea, Davis Strait, Baffin Bay, Denmark Strait, Iceland, Jan Mayen, the Greenland Sea, and Svalbard.
    \item CARRA-EAST, which includes Svalbard, Jan Mayen, Franz Josef Land, Novaya Zemlya, Barents Sea, and the northern regions of the Norwegian Sea and Scandinavia.
\end{itemize}

These domains encompass the Arctic’s four largest land ice bodies, including the Greenland Ice Sheet and the Austfonna Ice Cap, which are focal points in climate research. CARRA’s high resolution allows for accurate representation of critical Arctic-specific weather phenomena, such as polar lows in the North Atlantic and Greenland’s downslope katabatic winds, which have significant implications for local communities and maritime activities.

% \begin{figure}[htbp]
%     \centering
%     \includegraphics[width=0.33\textwidth]{northern_domain.png}
%     \caption{Domain of CARRA-EAST}
%     \label{fig:domain-CARRA}
% \end{figure}

The CARRA reanalysis is driven by ERA5 global reanalysis data for boundary conditions, integrating enhanced observations and ArcticDEM-based elevation data. CARRA also provides critical improvements in data quality through the assimilation of satellite observations (e.g., radiances, scatterometer data) and refined upper-air data algorithms, making it a powerful tool for Arctic climate monitoring. Covering the period from September 1990 to the present, CARRA supports comprehensive climate analysis with monthly updates that maintain a 2-3 month latency, ensuring the dataset remains relevant and accurate for ongoing climate research and operational needs. 

For our works we will focus on the CARRA-EAST domain. 

\subsection{CERRA-EDA}\label{subsec:CERRA_EDA}
CERRA-EDA \cite{Ridal2024} is an Ensemble of Data Assimilation (EDA) system comprising 10 members. It is based on the same code as the deterministic CERRA system and uses the same observational data, with the exception of the Microwave Sounding Unit and has a lower 11-km horizontal resolution. 

Each ensemble member in CERRA-EDA is generated by perturbing observations prior to data assimilation. One control member is initialized without any perturbations, while the other nine members incorporate Gaussian-distributed random perturbations within the estimated observational errors. No perturbations are applied to Sea Surface Temperature (SST), sea ice, or physical parameters. The lateral boundary conditions for CERRA-EDA are provided by the ERA5 ensemble, which includes 10 members and operates with a 63-km horizontal grid.

The analyses generated by CERRA-EDA are used to produce six-hour forecasts, from which background error covariances are estimated. CERRA relies on CERRA-EDA’s output, which supplies essential background error statistics and is integral to the data assimilation process. 

CERRA-EDA is run in advance of the deterministic system, though this typically does not hinder production due to CERRA-EDA’s faster processing enabled by its lower resolution and longer time steps, with the 10 ensemble members operating in parallel.

\section{Methodology}\label{sec:Methodology}
In this section we quickly review our signal and noise scheduling methodology before delving into the mathematical reasoning behind the control of variance via the number of diffusion steps. In particular, we derive an explicit formula to predict the variance of the generated data, which shows how it depends on the number of diffusion steps.

\subsection{Signal and noise rates}\label{subsec:signal_noise_rates}

\begin{figure} [htbp]
    \centering
    \includegraphics[width=0.9\textwidth]{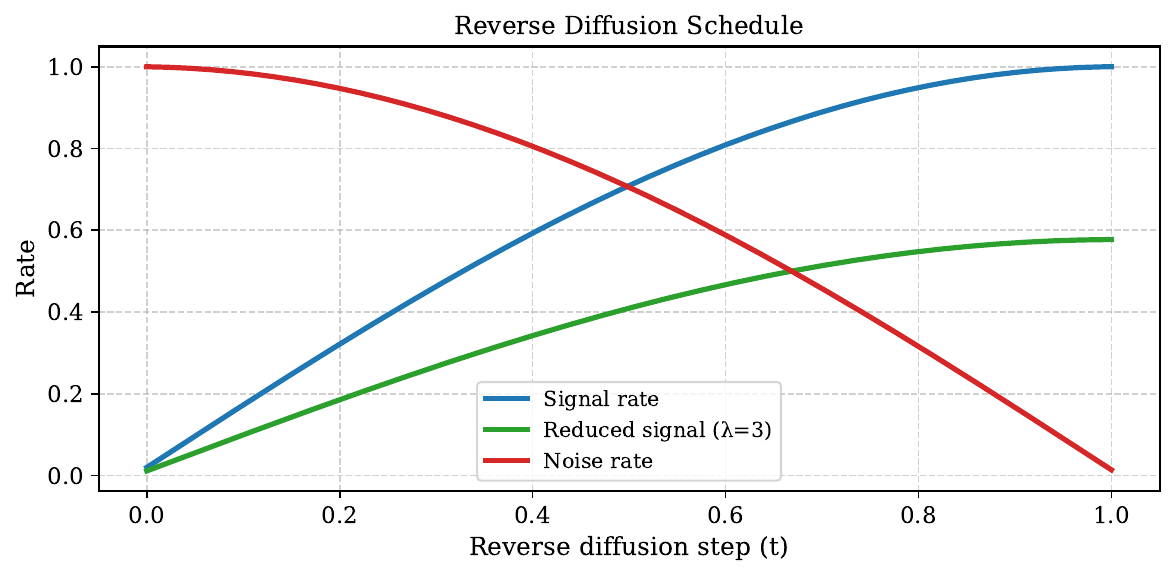}
    \caption{Reverse diffusion signal and noise ratios with cosine scheduling, including the reduced signal rate.}
    \label{fig:reducing factor}
\end{figure}

%Diffusion models operate by gradually corrupting and reconstructing data. During training, a random signal-to-noise ratio (SNR) is applied to the data, and the model is trained to remove the noise while being informed of its magnitude. During sampling, or reverse diffusion, the process starts with random noise and progressively reconstructs the data by decreasing the noise rate and increasing the signal rate at each step.

%In reverse diffusion, time steps tt are sampled uniformly from the range [1,0], divided into a set number of diffusion steps. Each step corresponds to a specific SNR, ensuring a smooth transition from noise-dominated inputs to fully reconstructed outputs.
In Section \ref{sec:Introduction} we discussed the importance of the diffusion schedule $\{ \alpha_t \}_{t = 1, \dots, T}$ in the generation process of DDIM.
Recently, various schedules have been proposed to find a perfect balance between the signal and noise in the diffusion process, including linear schedules \cite{ho2020denoising} and sinusoidal schedules \cite{nichol2021improved, song2022denoisingdiffusionimplicitmodels}. In this study, we adopt sinusoidal functions due to their ability to achieve smooth and controlled transitions between signal and noise components, which is crucial for effective denoising and high-quality image generation. In particular, we consider the following choice for the signal rate as in \cite{asperti2023image}:
\begin{align}\label{eq:scheduling_1}
    \sqrt{\alpha_t} = \sin \left(\pi \frac{T - t}{T} \right).
\end{align}
Therefore, the noise rate becomes:
\begin{align}\label{eq:scheduling_2}
    \sqrt{1 - \alpha_t} = \cos \left(\pi \frac{T - t}{T}\right), 
\end{align}
for which it holds that $\left( \sqrt{\alpha_t} \right)^2 + \left( \sqrt{1 - \alpha_t} \right)^2 = \sin \left(\pi \frac{T - t}{T}\right)^2 + \cos \left(\pi \frac{T - t}{T}\right)^2 = 1$, ensuring that the noisy images, formed as a weighted sum of signal and noise, consistently maintain unit variance \cite{nichol2021improved,Chen2023}.
The extremes of the schedule (i.e. when $t = 0$ and $t = T$) can be problematic, as near-pure signal or noise levels often cause instability. Following common practice \cite{Chen2023,Hoogeboom2023}, we clamp the extremes to maintain a balanced Signal-to-Noise Ratio (SNR), preventing degenerate cases and ensuring stable training and reliable reverse diffusion.

% \begin{align}
% \left\{
% \begin{aligned}
% \sqrt{1-\alpha_t} &= \sin(t) \\
% \sqrt{\alpha_t} &= \cos(t)
% \end{aligned}
% \right.
% \label{eq:scheduling}
% \end{align}
%Our scheduling approach discretizes the range into uniform steps, where each step $t$ determines the noise rate as $sin(t)$ and the signal rate as $cos(t)$, as shown in Equation \ref{eq:scheduling}. These rates satisfy $sin^2(t) + cos^2(t) = 1$, ensuring that the noisy images, formed as a weighted sum of signal and noise, consistently maintain unit variance \cite{nichol2021improved,Chen2023}.

As for the time discretization, we consider a uniform subdivision of the $[0, T]$ range into $T$ intervals, so that $t \in \{ 0, 1, \dots, T\}$. Note that, when a step-size $\Delta t > 1$ is considered, then clearly the time-domain becomes:
\begin{align}
    t \in \{ 0, \Delta t, 2\Delta t, \dots, N \Delta t \},
\end{align}
where $N = \frac{T}{\Delta t}$ represents the number of timesteps of our generative process.

\subsection{Controlling the variance}\label{subsec:controlling_variance}
The generative process in DDIM consists of a sequence of denoising operations alternated with the re-injection of the same noise. Our findings reveal that the key factor in controlling the variance lies in the choice of the number of diffusion steps. In this section, we analyze the evolution of the variance throughout the iterative diffusion process and provide an intuition of how the number of diffusion steps $N$ influences the variance of the generated data. Recall that the diffusion process in DDIM is defined by sampling $\x_0 \sim \mathcal{N}(\boldsymbol{0}, \boldsymbol{I})$ and then iterating through:
\begin{align}\label{eq:iteration}
    \x_t = \sqrt{\frac{\alpha_t}{\alpha_{t-\Delta t}}}\x_{t-\Delta t} + c_{t-\Delta t} \boldsymbol{\epsilon}_\Theta^{t-\Delta t}(\x_{t-\Delta t}),
\end{align}
where we define $c_{t-\Delta t}$ as:
\begin{align}\label{eq:c_epsilon_definition}
    c_{t - \Delta t} := \sqrt{1 - \alpha_t} - \sqrt{\frac{\alpha_t(1 - \alpha_{t-\Delta t})}{\alpha_{t-\Delta t}}},
\end{align}
to simplify the notation. The diffusion schedule $\{ \alpha_t \}_{t = 0, \dots, T}$ is chosen such that $\alpha_0 \approx 1$ and $\alpha_T \approx 0$. Note that $\alpha_T \neq 0$, as its value is clamped to avoid instabilities during training, as already discussed in \cref{subsec:signal_noise_rates}. 

{Let $\x_0$ be an $n$-dimensional random variable with Gaussian distribution such that $\mathbb{E} [ \x_0 ] = \boldsymbol{0}$ and $Var(\x_0) = \boldsymbol{I}$, and let $\{ \x_t \}_{t = 0, \dots, T}$ be the stochastic process defined in \eqref{eq:iteration}. 
Since the full covariance matrix $Var(\x_t)$ has dimension $n \times n$, its computation is prohibitive for high-dimensional data such as images. 
Therefore, in this work we restrict our analysis to the \emph{total variance}, defined as the sum of the diagonal elements of the covariance matrix:
\begin{align}\label{eq:totVar}
    \TotVar(\x_t) := \Tr \!\big( Var(\x_t) \big)
    = \sum_{i=1}^{n} Var(\x_t)_{i,i}.
\end{align}
The total variance provides a scalar measure of the overall uncertainty of $\x_t$, summarizing how much the process spreads around its mean across all pixels while ignoring pairwise correlations. 
This simplification allows us to capture the global evolution of uncertainty along the diffusion trajectory, which is sufficient for our goal of analyzing the overall stability and noise amplification of DDIM downscaling, without delving into spatial dependencies between adjacent pixels. In the following, we report for simplicity the results for the scalar case (i.e., $n = 1$), and we leave the complete proofs for the general $n$-dimensional case to the appendix.}

{The theoretical derivation describing how the variance evolves through the diffusion process requires estimating the variance $\TotVar_t$ at the $t$-th step based on $\TotVar_{t-\Delta t}$ from the previous step. The following Proposition establishes a linear relationship between the two quantities.}

{\begin{proposition}\label{prop:recursive_formula}
    Let $\boldsymbol{m}_t:= \mathbb{E}[\x_t]$ and $\v_t := \ewVar(\x_t)$. Then, for any $t = \Delta t, \dots, N \Delta t$, it holds:
    \begin{align}\label{eq:recursive_formula}
        \v_t \approx \boldsymbol{F}_t \v_{t-\Delta t} + \boldsymbol{g}_t,
    \end{align}
    where:
    \begin{align*}
        &\boldsymbol{F}_t := \left(\sqrt{\frac{\alpha_t}{\alpha_{t - \Delta t}}}  + c_{t - \Delta t}\boldsymbol{J}_{t - \Delta t} \right)^2, \\
        &\boldsymbol{g}_t := c_{t - \Delta t}^2 \ewVar(\boldsymbol{r}_t),
    \end{align*}
    and $\boldsymbol{r}_t$ is a random variable independent on $\x_t$.
\end{proposition}}

{\begin{proof}
    Linearizing $\boldsymbol{\epsilon}_\Theta^{t - \Delta t} (\x_{t - \Delta t})$ around $\boldsymbol{m}_{t - \Delta t} := \mathbb{E} \left[ \x_{t - \Delta t} \right]$ via the first order Taylor decomposition yelds:
    \begin{align}
        \boldsymbol{\epsilon}_\Theta^{t - \Delta t} (\x_{t - \Delta t}) \approx \boldsymbol{E}_{t - \Delta t} + \boldsymbol{J}_{t - \Delta t} \left( \x_{t - \Delta t} - \boldsymbol{m}_{t - \Delta t} \right) + \boldsymbol{r}_t,
    \end{align}
    where we defined $\boldsymbol{E}_{t-\Delta t} := \boldsymbol{\epsilon}_\Theta^{t - \Delta t} (\boldsymbol{m}_{t - \Delta t})$, $\boldsymbol{J}_{t-\Delta t} := \nabla_{\boldsymbol{m}_{t - \Delta t}} \boldsymbol{\epsilon}_\Theta^{t - \Delta t}(\boldsymbol{m}_{t - \Delta t})$, and $\boldsymbol{r}_t$ is a remainder term, independent on $\x_{t-\Delta t}$. Plugging this expression into \eqref{eq:generative_process_timestep} leads to
    \begin{align*}
    	\x_t  &= \sqrt{\frac{\alpha_t}{\alpha_{t - \Delta t}}} \x_{t - \Delta t} + c_{t - \Delta t} \boldsymbol{\epsilon}_\Theta^{t - \Delta t} (\x_{t - \Delta t}) \\ & \approx \sqrt{\frac{\alpha_t}{\alpha_{t - \Delta t}}} \x_{t - \Delta t} + c_{t - \Delta t} \left(\boldsymbol{E}_{t-\Delta t}+ \boldsymbol{J}_{t - \Delta t} \left( \x_{t - \Delta t} - \boldsymbol{m}_{t - \Delta t} \right) + \boldsymbol{r}_t \right) \\ & = \left( \sqrt{\frac{\alpha_t}{\alpha_{t - \Delta t}}} + c_{t - \Delta t}\boldsymbol{J}_{t - \Delta t} \right) \x_{t - \Delta t} + c_{t - \Delta t} \boldsymbol{r}_t + c_{t - \Delta t} \boldsymbol{E}_{t-\Delta t} - c_{t - \Delta t}\boldsymbol{m}_{t - \Delta t} \boldsymbol{J}_{t - \Delta t}.
    \end{align*}
    Computing the variance on both side of the above equation, using the bilinearity of the variance, and ignoring terms that are constant in $\x_{t-1}$ as their variance is zero, yields:
    \begin{align*}
    	\ewVar(\x_t) \approx \left(\sqrt{\frac{\alpha_t}{\alpha_{t - \Delta t}}}  + c_{t - \Delta t}\boldsymbol{J}_{t - \Delta t} \right)^2 \ewVar(\x_{t - \Delta t}) + c_{t - \Delta t}^2 \ewVar(\boldsymbol{r}_t),
    \end{align*}
    which corresponds to \eqref{eq:recursive_formula} by setting $\boldsymbol{F}_t := \left(\sqrt{\frac{\alpha_t}{\alpha_{t - \Delta t}}}  + c_{t - \Delta t}\boldsymbol{J}_{t - \Delta t} \right)^2$ and $\boldsymbol{g}_t := c_{t - \Delta t}^2 \ewVar(\boldsymbol{r}_t)$. 
\end{proof}}

{Equation \eqref{eq:recursive_formula} represents a first-order linear recurrence relation of dimension $n$ with non-constant coefficients, for which a closed-form expression can be derived to compute $\v_T$, the variance of the generated data, as summarized in the following Theorem.} \\

{\begin{theorem}\label{teo:closed_formula}
    Let $\boldsymbol{F}_t$ and $\boldsymbol{g}_t$ defined as in Proposition \ref{prop:recursive_formula}, and let $\Delta t$ be a fixed step-size for DDIM, so that any timestep $t$ can be written as $i \Delta t$ for $i = 0, \dots, N$. Note that $\v_0 = 1$, as $\x_0 \sim \mathcal{N}(0, 1)$. Then, defining $\boldsymbol{F}_{(N+1)\Delta t} = 1$ to simplify the notation, it holds:
    \begin{align}\label{eq:closed_formula}
        \v_T \approx \left(\prod_{i = 1}^N \boldsymbol{F}_{i \Delta t} \right) + \sum_{i = 1}^N \left[ \left( \prod_{k = i+1}^{N+1} \boldsymbol{F}_{k \Delta t} \right) \boldsymbol{g}_{i \Delta t}\right].
    \end{align}
\end{theorem}
\begin{proof}
    The proof of this Theorem is mainly a recursive application of Proposition \ref{prop:recursive_formula} to compute $\v_T$ from $\v_0$. Indeed, we recall that:
    \begin{align*}
        \v_t \approx \boldsymbol{F}_t \v_{t - \Delta t} + \boldsymbol{g}_t.
    \end{align*}
    Since, as already observed, $\v_0 = 1$, then:
    \begin{align*}
        &\v_{\Delta t} \approx \boldsymbol{F}_{\Delta t}  + \boldsymbol{g}_{\Delta t}, \\
        &\v_{2\Delta t} \approx \boldsymbol{F}_{2\Delta t} \boldsymbol{F}_{\Delta t}  + \boldsymbol{F}_{2 \Delta t}\boldsymbol{g}_{\Delta t} + \boldsymbol{g}_{2\Delta t}, \\
        &\v_{3\Delta t} \approx \boldsymbol{F}_{3\Delta t}\boldsymbol{F}_{2\Delta t} \boldsymbol{F}_{\Delta t}  + \boldsymbol{F}_{3\Delta t}\boldsymbol{F}_{2 \Delta t}\boldsymbol{g}_{\Delta t} + \boldsymbol{F}_{3\Delta t}\boldsymbol{g}_{2\Delta t} + \boldsymbol{g}_{3 \Delta t}, \\
        & \dots \\
        & \v_{N \Delta t} \approx \left( \prod_{i = 1}^N \boldsymbol{F}_{i\Delta t} \right)  + \sum_{i=1}^N \left( \prod_{k=i+1}^{N+1} \boldsymbol{F}_{k \Delta t} \right) \boldsymbol{g}_{i \Delta t},
    \end{align*}
    concluding the proof.
\end{proof}}

{Notably, \Cref{teo:closed_formula} not only enables the explicit computation of the variance $\v_T$, but also highlights its dependence on the number of diffusion timesteps. A key result in this regard is presented in the following Theorem.}

{\begin{theorem}\label{teo:monotonicity}
    Let $\boldsymbol{F}_t$ and $\boldsymbol{g}_t$ defined as in Proposition \ref{prop:recursive_formula}, and let $\Delta t$ be the discrete step-size for DDIM, so that any timestep $t$ can be written as $i \Delta t$ for $i = 0, \dots, N$, and $N = \frac{T}{\Delta t}$ Then $\{ \v_{N \Delta t} \}_{N \in \mathbb{N}}$ is a monotonically increasing sequence and:
    \begin{align}\label{eq:limit_variance}
        \lim_{N \to \infty} \v_{N \Delta t} =  e^{- \int_0^T \lambda(s) ds} + \int_0^T e^{- \int_0^s \lambda(\kappa)d\kappa} \gamma(s) ds,
    \end{align}
    where:
    \begin{align*}
    \begin{split}
        &\lambda(t) = - \partial_t \log \alpha(t - \Delta t) + \boldsymbol{J}_{t - \Delta t} \alpha'(t - \Delta t) \left( \frac{\sqrt{1 - \alpha(t - \Delta t)}}{\alpha(t - \Delta t)} + \frac{1}{\sqrt{1 - \alpha(t - \Delta t)}} \right), \\
        &\gamma(t) = \left(\frac{\alpha'(t - \Delta t)\sqrt{1 - \alpha(t - \Delta t)}}{2\alpha(t - \Delta t)} +\frac{\alpha'(t - \Delta t)}{2\sqrt{1 - \alpha(t - \Delta t)}}\right) \TotVar(\boldsymbol{r}_t)
    \end{split}
    \end{align*}
\end{theorem}
\begin{proof}[Sketch of the proof]
    We report here only the main ideas of the derivation, while the complete proof is provided in Appendix~\ref{app:scalar_proof}. A first-order Taylor expansion of $\sqrt{\alpha(t)}$ and $\sqrt{1-\alpha(t)}$ around $t-\Delta t$
    shows that:
    \begin{align*}
        &\boldsymbol{F}_t = 1 + \lambda_{t-\Delta t} \Delta t + O(\Delta t^2) \\
        &\boldsymbol{g}_t = \gamma_{t-\Delta t} \Delta t + O(\Delta t^2),
    \end{align*}
    where $\lambda_{t - \Delta t}$ and $\gamma_{t - \Delta t}$ depend on $\alpha'(t - \Delta t)$ and the local Jacobian $\boldsymbol{J}_{t - \Delta t}$. \\
    Substituting in the expression of Theorem \ref{teo:closed_formula} leads to a product–sum expression whose discrete factors $\prod_{k=1}^T \boldsymbol{F}_k$ converge monotonically from below to $\exp\!\big(\int_0^T \lambda(s)\,ds\big)$ as $\Delta t \to 0$.  
    The variance sequence $\{ \TotVar_{N\Delta t} \}_{N \in \mathbb{N}}$ therefore satisfies
    $$
        \TotVar_{N\Delta t} 
        \overset{N \to \infty}{\longrightarrow}  e^{-\int_0^T \lambda(s)\,ds}
          + \int_0^T e^{-\int_0^s \lambda(\kappa)\,d\kappa}\,\gamma(s)\,ds,
    $$
    completing the proof.
\end{proof}}

{Theorem~\ref{teo:monotonicity} reveals a compelling property regarding the relationship between the variance of the diffusion process at the final step, $\TotVar_T$, and the number of diffusion steps performed. Specifically, it shows that the variance increases monotonically with the number of steps $N$, demonstrating that it is possible to calibrate the variance of the generated data by adjusting the number of diffusion steps, as further discussed in the next section. Moreover, \eqref{eq:limit_variance} indicates that $\TotVar_T$ cannot grow indefinitely, but instead converges to a finite limit determined by the schedule $\{ \alpha_t \}_{t \in [0, T]}$ and by the neural network $\epsilon_\Theta^t$ through the Jacobian $\boldsymbol{J}_t$. To the best of our knowledge, this is the first work that theoretically proves that the variance expressed by diffusion models can be calibrated by tuning the number of diffusion steps. In the following section, we empirically demonstrate that this property holds in practice by testing variance calibration on conditioned diffusion models applied to ERA5-to-CERRA downscaling.}

\section{Experimental Setting}
In this section, we describe the pre-processing, training, and evaluation strategies required to adapt the diffusion model for high-resolution downscaling across the full CERRA domain. The transition to larger spatial scales necessitated adjustments to both data handling and model architecture to ensure consistent performance. We focus on data selection and pre-processing in Section \ref{subsec:data_preprocess}, followed by a detailed explanation of the modifications required for training on the full domain in Section \ref{subsec:fulldomain_training}.

\subsection{Data selection and pre-processing}\label{subsec:data_preprocess}
The diffusion model performs downscaling in an image-to-image manner, transforming low-resolution ERA5 data into high-resolution CERRA or CARRA outputs while maintaining alignment over the same spatial domain for input and output. The primary experimental framework of our study focuses on downscaling wind speed from ERA5 to CERRA over the full CERRA domain and comparing the ensemble results with the CERRA-EDA dataset. Our model will be trained solely on the CERRA data and no input of CERRA-EDA is used during training, ensuring that the model will learn the ensemble variability only from the CERRA data itself. 

The CERRA domain spans a grid of 1069×1069 points across Europe, represented in a Lambert Conformal Conic (LCC) projection. The complete CERRA dataset covers the period from September 1984 to June 2021; for our experiments, we utilized data from 1985 to 2010 for training and from 2016 for ensemble testing. Additionally, we evaluated overall performance for the entire decade from 2011 to 2020, though computational constraints limited ensemble variance testing to the year 2016.
In our diffusion model, ERA5 data serves as the conditioning input, re-gridded to align with CERRA’s LCC grid. Wind speed, $s$, is derived from the two wind components, $s_u$ and $s_v$, as:
\begin{align*}
    s := \left\| \begin{bmatrix}
        s_u \\ s_v
    \end{bmatrix} \right\|_2 = \sqrt{s_u^2 + s_v^2}
\end{align*}

Our diffusion model is based on a residual convolutional denoiser, which benefits from input dimensions divisible by two at multiple levels. To ensure consistency between the downsample and upsample paths, we augment the original grid size (1069×1069) with mirroring padding, resulting in a final input size of 1072×1072. Our conditioning is implemented via concatenation on the input of the denoising network, to equalize the dimension we opt to upscale the ERA5 data to $1069 \times 1069$ via bilinear interpolation, and then we apply the same mirroring padding to ensure spatial consistency. 

To compare our diffusion-based ensembles with the existing CERRA-EDA ensemble members, we note that CERRA-EDA operates at half the spatial resolution of CERRA (565×565) and a temporal resolution of six hours, also halved relative to CERRA. For consistency, we train our model exclusively on full-resolution CERRA data with a three-hour temporal resolution. Wind speed is an instantaneous value, therefore six-hourly outputs can be generated using conditioning information at the same temporal resolution. The generated outputs retain the full CERRA resolution; for comparison with CERRA-EDA, we apply bilinear interpolation to downscale the results.

Diffusion models require standardized inputs to balance signal and noise rates effectively. We apply a standardizer that normalizes inputs using the mean and variance computed across the training set, ensuring consistent scaling across batches. It is important to note that this variance is relative to the training data and is distinct from the ensemble variance between members produced by the diffusion model. This normalization is applied to both high-resolution images (signal rates) and low-resolution conditioning data, aligning their scales and simplifying the learning task. Experimentally, this approach demonstrated improved performance and stability.

\subsection{Training on the full domain}\label{subsec:fulldomain_training}

The general structure of our diffusion model follows the framework established in \cite{Merizzi2024}. The denoiser is a residual U-Net \cite{ronneberger2015,7780459} implemented in TensorFlow/Keras, designed as a symmetric encoder–decoder network with skip connections and residual convolutional blocks using Swish activations and layer normalization. The architecture performs three stages of downsampling through average pooling and three corresponding upsampling stages via bilinear interpolation. Each stage (both in the encoder, decoder, and latent bottleneck) contains three residual blocks, each composed of two 3×3 convolutions and a skip connection to preserve spatial detail. The model contains approximately 22 million learnable parameters, predicting the noise component at each diffusion step.

Training is conducted by uniformly sampling signal-to-noise ratios from the sinusoidal noise scheduling functions, optimizing the mean absolute error (MAE) between the sampled noise and the predicted noise.

Scaling the diffusion model to the full CERRA domain, with image sizes of 1069×1069, required modifications to a model originally designed for downscaling images up to 256×256. These adaptations involved adjustments to both the noise schedule and the network architecture to ensure effective performance at higher resolutions.

Recent findings \cite{Chen2023, Hoogeboom2023} have demonstrated that cosine noise schedules can reduce the effective noise added by the model as image size increases. This occurs because redundancy in data, such as correlations among nearby pixels, typically increases with resolution, and the independently added noise becomes easier to denoise in larger images. Consequently, noise schedules optimized for smaller resolutions may not perform as effectively at higher resolutions. To address this issue, we introduce a constant scaling factor $\lambda > 1$ to reduce the signal rate in the noise schedule. For an input $\x_T \sim p_{gt}(\x_T)$ and a scaling factor $\lambda$, the noise schedule \cref{eq:forward_process} is modified as follows: 

\begin{align}\label{eq:reduce_factor}
\x_t = \sqrt{\alpha_t} \frac{\x_T}{\lambda} + \sqrt{1 - \alpha_t}\boldsymbol{\epsilon}_t 
\end{align}

Experimentally, setting the value $\lambda = 3$ improved training efficiency and overall model performance, which is a crucial aspect for large-scale ensemble diffusion applications. A comparison of signal and noise ratios with the scaled signal rate is shown in \Cref{fig:reducing factor}. Notably, applying the same scaling to the low-resolution conditioning signal further stabilized the training process.

Experimentally, we tested several values of $\lambda \in [1,4]$ and observed that $\lambda = 3$ provided the best trade-off between reconstruction fidelity and training stability, achieving the lowest MSE with respect to the deterministic CERRA ground truth. Full ablation at multiple $\lambda$ values and full resolution would be computationally prohibitive; therefore, smaller-scale experiments were used to guide the choice, keeping the spatial resolution constant as it is the key factor influencing this effect. The selected value should thus be regarded as empirically determined. This adjustment proved crucial for large-scale ensemble diffusion applications, improving convergence and overall stability. Notably, applying the same scaling to the low-resolution conditioning signal further stabilized the training process.

While the core U-Net architecture remained largely unchanged, scaling to larger domains required increasing the number of residual blocks in the bottleneck. This adjustment improved performance when generating larger images \cite{Hoogeboom2023}. The model’s performance across the increased domain remained comparable to that observed at smaller resolutions, with the bottleneck modification enhancing its ability to process higher-resolution inputs effectively.

The diffusion model is trained on the ERA5 to CERRA task for wind speed on the full domain, utilizing data from 1985 to 2010 for training. The training was performed on a single compute node equipped with three NVIDIA A100 GPUs (64~GB VRAM each), using a \textit{batch separation (mirroring) strategy} to distribute the workload across devices. The model was trained for 400~epochs with 500~steps per epoch, corresponding to a total training time of approximately 42~hours. Each epoch required about 300~seconds to complete, and the batch size was set to three, with one sample processed per GPU. \textit{AdamW} was used as the optimizer, with an initial training phase of 200~epochs using a learning rate of $1\times10^{-4}$ and a weight decay of $1\times10^{-5}$, followed by a fine-tuning phase of 100~epochs with a reduced learning rate of $1\times10^{-5}$ and a weight decay of $1\times10^{-6}$. The model was trained to minimize the mean absolute error (MAE) between the predicted and true noise. During inference, the memory constraints are relaxed, and the model can be executed on GPUs with as little as 16~GB of VRAM.

\subsection{Implementation Details}

The denoiser is a residual U-Net implemented in TensorFlow/Keras, featuring symmetric encoder–decoder paths with skip connections and residual convolutional blocks using Swish activations and layer normalization. Downsampling is done via average pooling, upsampling via bilinear interpolation, and the sinusoidal embedding of the diffusion timestep is concatenated to the input features

Please add model size, training duration, GPU hours, and inference latency vs. steps $N$ on the full grid (1069×1069), plus memory use. This will help weigh calibration quality against runtime; your hardware/optimizer details are noted but not quantified.

\section{Experiments}\label{sec:Experiments}

Our primary experimental objective is to evaluate the behavior of ensemble variance while changing the timestep $\Delta t$. While a comprehensive analysis of the ensemble diffusion model's performance across the entire CERRA domain is beyond the scope of this paper, we ensure that the model operates within a realm of satisfactory performance. To achieve this, we begin—similar to \cite{Merizzi2024}—by comparing our model's performance against bilinearly interpolated ERA5 data and a residual U-Net model trained for the same task. The results, obtained by computing the Mean Squared Error (MSE) and SSIM between the high-resolution output produced by the model and the original CERRA, are shown in \Cref{tab:metrics_comparison}.

\begin{table*}[htbp]
\centering
\def\arraystretch{1} % for increased vertical spacing
\begin{tabular}{>{\bfseries}lccc}
    \toprule
    & \textbf{Bilinear} & \textbf{U-Net} & \textbf{DDIM (Our)} \\
    \midrule
    \textbf{MSE} $(\downarrow)$ & $3.73\text{e-03}$ & $2.50\text{e-04}$ & $2.54\text{e-04}$ \\
    \textbf{SSIM} $(\uparrow)$& $0.751$ & $0.892$ & $0.923$ \\
    \bottomrule
\end{tabular}
\caption{Performance of ensemble diffusion against bilinear interpolation and U-Net for the years 2011-2020.}
\label{tab:metrics_comparison}
\end{table*}
Testing for the years 2011–2020, the ensemble diffusion model achieves an MSE of $2.54 \cdot 10^{-4}$, compared to $2.50 \cdot 10^{-4}$ for the U-Net and $3.73 \cdot 10^{-3}$ for bilinear interpolation. For SSIM, bilinear interpolation scores 0.751, the U-Net scores 0.892, and ensemble diffusion achieves the highest score of 0.923. Ensemble diffusion was executed with $N = 2$ diffusion steps and an ensemble of 10 members. 

These results confirm that the diffusion model delivers satisfactory performance, with MSE values approximately an order of magnitude better than bilinear interpolation and comparable to those of the U-Net. Additionally, as expected, the diffusion model outperforms both alternatives in terms of SSIM. The proposed metrics were calculated on normalized data. 

\subsection{Quantifying variance}

\begin{figure}
    \centering
    \includegraphics[width=\textwidth]{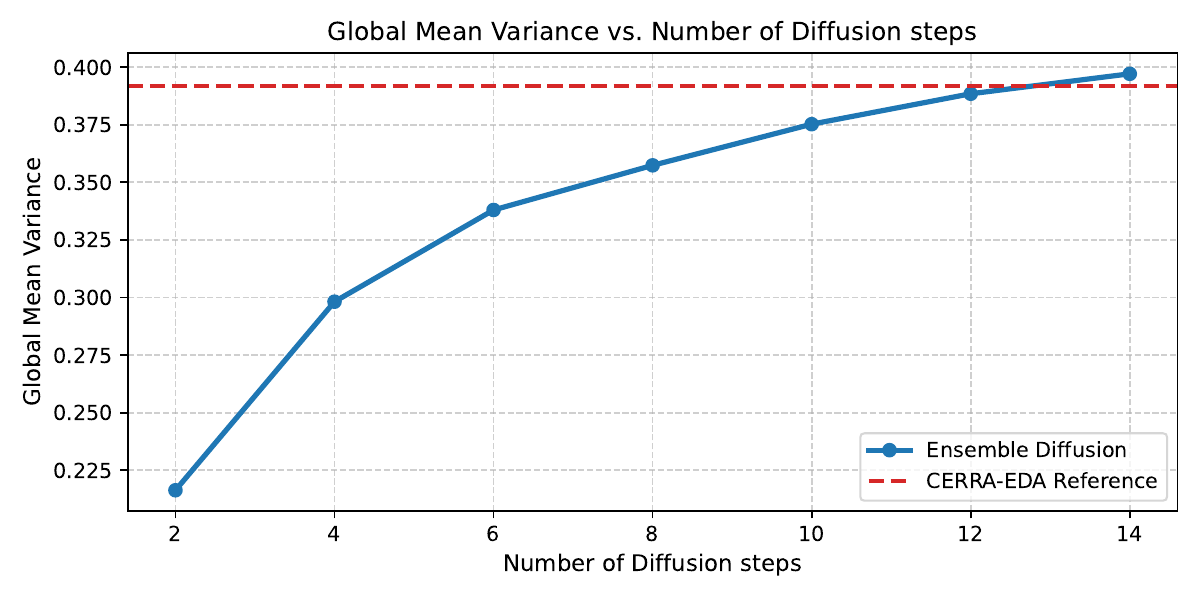}
    \caption{Plotting the global mean variance $\mu_V$ of ensemble diffusion for different number of steps against CERRA-EDA.}
    \label{fig:global_variance_graph}
\end{figure}

The first aspect of variance analysis we evaluate in our experiments is understanding how the model's expressed variance changes with the number of diffusion steps, as detailed in Section \ref{subsec:controlling_variance}. For this experiment we selected the year 2016 and computed the ensemble variance for 10 members for all the year at 6h intervals, matching the existing format of the CERRA-EDA ensemble. Furthermore, we used bilinear interpolation to reduce the spatial dimension of our output image from $1069 \times 1069$ to $565 \times 565$, also matching the existing CERRA-EDA format. Having obtained a comparable dataset, we proceeded to calculate the pixel-wise variance for the ensemble member of each ensemble of the year, finally calculating the mean variance, a value representing the overall variance expressed by the diffusion model. 

Given a dataset of ensembles $\boldsymbol{D} \in \mathbb{R}^{S \times M \times h \times w}$, where $S$ represents the number of samples for 2016, $M = 10$ is the number of ensemble members for each datapoint, and $(h, w) = (565, 565)$ are the spatial dimensions, we compute the pixel-wise variance across the ensemble members $\boldsymbol{V} \in \mathbb{R}^{S \times M \times 565 \times 565}$, such that for each sample $i = 1, \dots, S$, and any spatial coordinate $(x, y) \in \{ 1, \dots, h \} \times \{ 1, \dots, w \}$,
\begin{align}\label{eq:pixelwise_variance}
    \boldsymbol{V}_i(x, y) = \frac{1}{M} \sum_{j=1}^M \left( \boldsymbol{D}_{i, j}(x, y) - \frac{1}{M} \sum_{k=1}^M \boldsymbol{D}_{i, k}(x, y) \right)^2.
\end{align}
We then compute the mean variance across all samples and spatial locations as:
\begin{align}\label{eq:mean_variance}
    \mu_{\boldsymbol{V}} = \frac{1}{S \cdot h \cdot w} \sum_{i=1}^{S} \sum_{x=1}^{h} \sum_{y=1}^{w} \boldsymbol{V}_i(x, y).
\end{align}
Here, $\mu_{\boldsymbol{V}}$ is a scalar value representing the global mean variance across all samples and pixels. By computing $\mu_{\boldsymbol{V}}$ for different diffusion steps, we observe that the global variance increases with the number of steps, as illustrated in Figure \ref{fig:global_variance_graph}. Furthermore, we demonstrate that the variance control mechanism successfully aligns the global variance with that of CERRA-EDA, with the closest match achieved at $N = 12$ diffusion steps.
Interestingly, the relationship between the number of steps and ensemble variance may have been overlooked in previous research, as it is common practice to run diffusion models with a high number of steps \cite{ling2024diffusion,price2023gencast, bassetti2024diffesm}. In such cases, the overall variance would have already converged to a fixed value, making it unaffected by changes in the number of reverse diffusion steps.

\subsection{Analyzing variance spatially}

\begin{figure*}[ht]
    \centering
    \includegraphics[width=\textwidth]{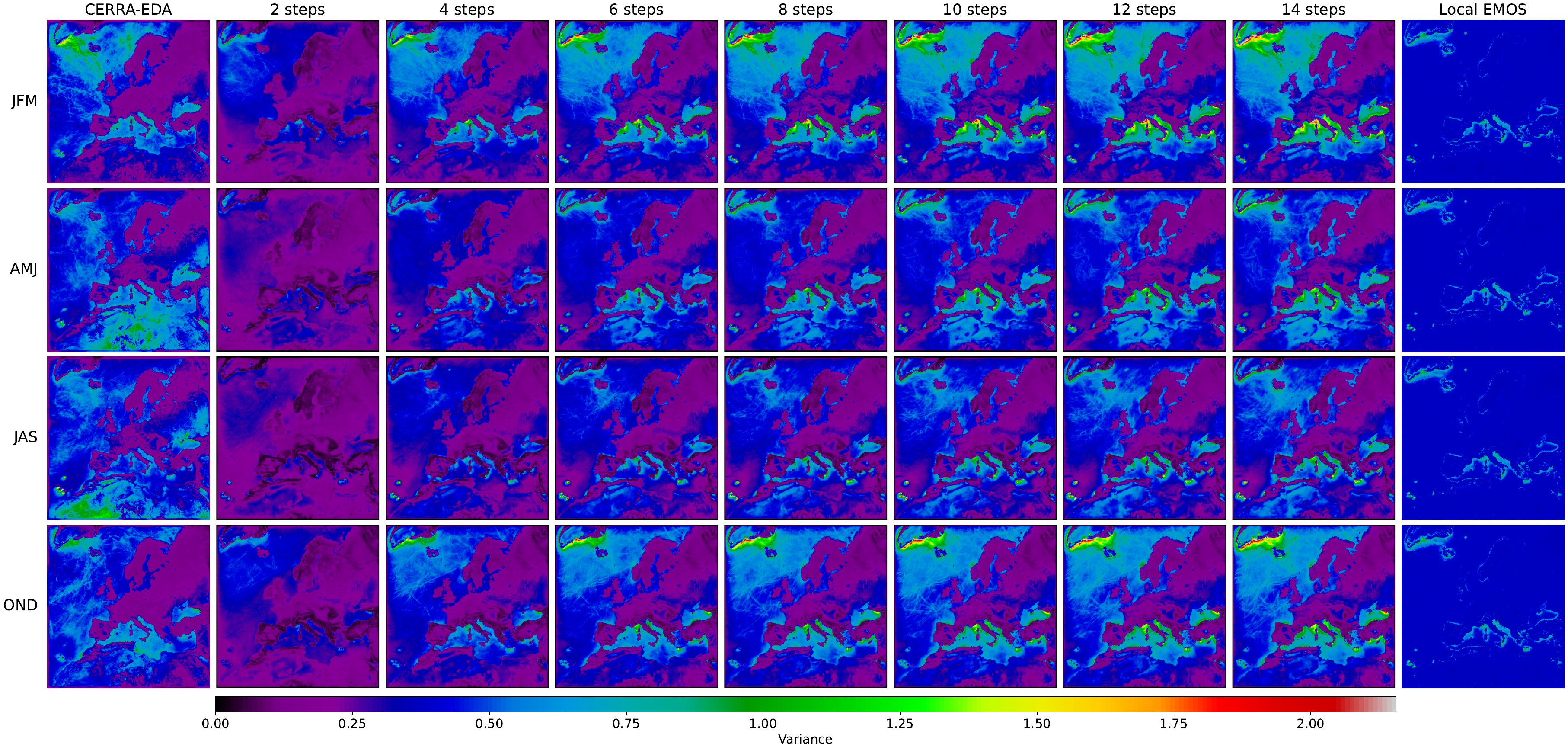}
    \caption{Comparing the spatial variance of CERRA-EDA with ensemble diffusion at different diffusion steps. The comparison relative to the testing year 2016 and is performed in a 3 monthly manner to highlight seasonal variations.}
    \label{fig:spatial_variance_big_picture}
\end{figure*}

The global mean-variance inform us that the diffusion model is able to match the overall amount of variance present in CERRA-EDA, but this gives us no information to where the variance is distributed in space. Our interest here is to examine if there is a match between the two spatial distributions of the variance. 

For the following experiments, we decided to investigate the variance in a three-monthly setting, to highlight the behavior of the model in different seasons. To achieve this goal we divided the the samples into four sets, including respectively January February and March (JFM), April May and June (AMJ), July August and September (JAS) and October November and December (OND). 

To analyze spatial variance, we compute the pixel-wise mean of variance maps across all samples within each three-monthly subset. Let $k \in \{JFM, AMJ, JAS, OND\}$ denote the season, $\mathcal{S}_k \subseteq \{ 1, \dots, S \}$ represent the subset of indices corresponding to the season $k$, and $S_k$ the number of samples corresponding to season $k$. The spatial mean variance $\mu^{spatial}_k$ for season $k$ is given by:

\begin{align}\label{eq:spatial_mean_variance}
    \boldsymbol{\mu}^{spatial}_k (x, y) = \frac{1}{S_k} \sum_{i \in \mathcal{S}_k} \boldsymbol{V}_i (x, y), \quad (x, y) \in \{ 1, \dots, h \} \times \{ 1, \dots, w \}.
\end{align}
Here, $\boldsymbol{V}_i$ and $\boldsymbol{\mu}^{spatial}_k$ are understood to represent spatial fields, where the value of each pixel is computed independently.
We compute the spatial mean-variance for the reference ensemble CERRA-EDA and for Ensemble Diffusion across a range of diffusion steps $N \in \{2, 4, 6, 8, 10, 12, 14 \}$ for the year 2016. This allows us to visualize how the spatial distribution of variance evolves as the number of diffusion steps increases. The obtained results are reported in Figure \ref{fig:spatial_variance_big_picture}. We observe substantial overlap between the variance distributions of CERRA-EDA and Ensemble Diffusion. This indicates that the diffusion model successfully captures the variability of the data by training solely on the CERRA dataset, without requiring additional information about the underlying uncertainty. Furthermore, we note that the degree of spatial overlap varies by season, with the highest overlap observed in the JFM and OND sets, while AMJ and JAS exhibit less overlap.

% However, a visual comparison alone is not sufficient to rigorously assess the overlap between the spatial mean-variance maps. 
% To numerically quantify this overlap, we compute the absolute error between the spatial mean-variance maps of Ensemble Diffusion and CERRA-EDA. Let $\mu_{k,\text{ED}}^{\text{spatial}}$ denote the spatial mean variance map of Ensemble Diffusion and $\mu_{k,\text{CE}}^{\text{spatial}}$ denote the one of CERRA-EDA. We thus define the mean-variance discrepancy score (MVD) for season $k$ as:

To numerically quantify the overlap between variance distributions, we introduce the mean-variance discrepancy (MVD) score, a metric defined as the absolute error between the spatial mean-variance maps for each season. Let $\boldsymbol{\mu}_{k,\text{ED}}^{\text{spatial}}$ denote the spatial mean-variance map of Ensemble Diffusion and $\boldsymbol{\mu}_{k,\text{CE}}^{\text{spatial}}$ denote the corresponding map for CERRA-EDA. The MVD score for season $k$ is then defined as:
\begin{align}\label{eq:mvd_score}
    \text{MVD}_k = \frac{1}{h \cdot w} \sum_{x = 1}^ h\sum_{y=1}^w \left| \boldsymbol{\mu}_{k,\text{ED}}^{\text{spatial}}(x, y) - \boldsymbol{\mu}_{k,\text{CE}}^{\text{spatial}}(x, y) \right|.
\end{align}
The numerical results for the MVD scores computed across all three-monthly periods are reported in Table \ref{tab:MVD_between_variance_images}. 

From the MVD results, we observe that the optimal yearly number of diffusion steps is $N=8$, a lower value if compared with 12, which was identified based on the global mean variance. This finding reveals that, while 12 steps yield an overall variance closest to the reference, the spatial distribution of the variance behaves differently. This underscores the importance of a tailored tuning procedure to achieve better alignment with expected uncertainties. Moreover, the MVD scores are not uniform throughout the year: the winter months achieve the lowest MVD scores with 4 diffusion steps, while the summer months perform best with 12 diffusion steps, possibly due to the model being affected by seasonal dynamics.

\begin{figure*}[ht]
    \centering
    \includegraphics[width=\textwidth]{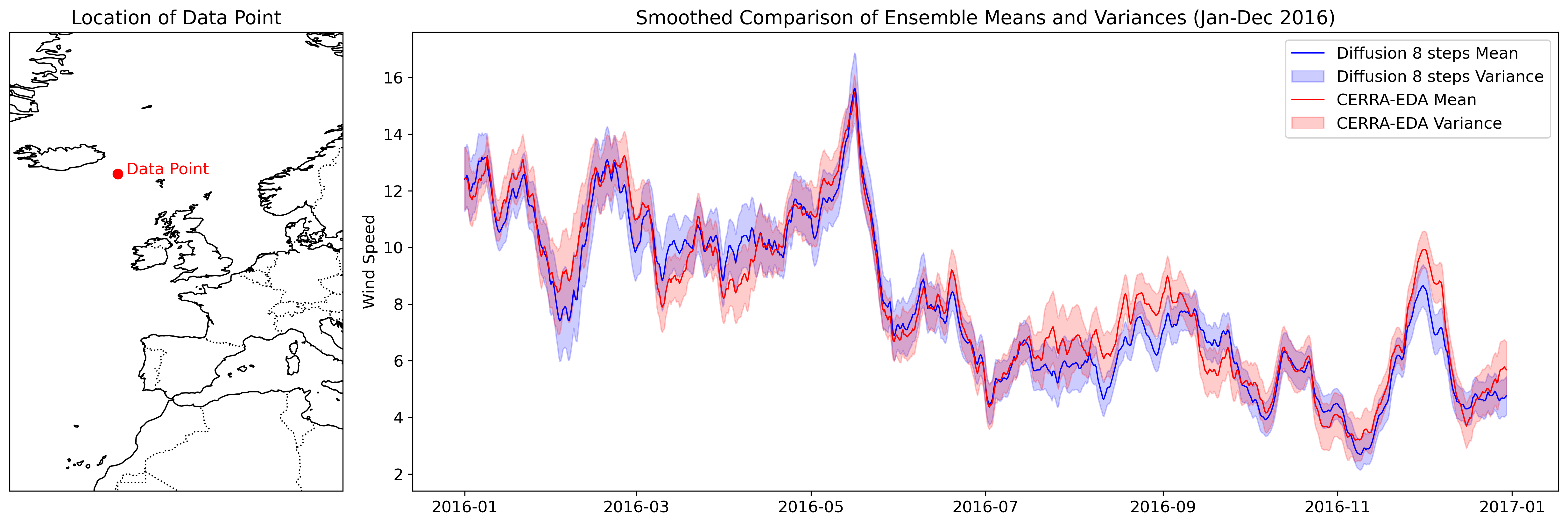}
    \caption{Mean and variance range at a randomly selected point in the domain, compared between ensemble diffusion (8 reverse diffusion steps) and CERRA-EDA. Data is smoothed using a 30 samples moving average window for readability and covers the entire year of 2016.}
    \label{fig:single_point_variance}
\end{figure*}

Alongside the full-domain analysis, we examine the behavior of the generated ensemble at a randomly selected point within the domain to provide a clearer and more readable visualization of the mean and variance ranges. For this analysis, we plot the mean value and variance range at the chosen point for both the 8-step ensemble diffusion and CERRA-EDA. To improve clarity over the high-frequency time series, we apply a moving average smoothing with a 30-sample window. The resulting plot, shown in Figure \ref{fig:single_point_variance}, reveals consistent overlap between the mean and variance ranges.

\begin{table*}[htbp]
\centering
\def\arraystretch{1.1} % for increased vertical spacing
\small % or \footnotesize to make it even more compact
\begin{tabular}{>{\bfseries}lcccccccc}
    \toprule
    \multicolumn{8}{c}{\textbf{MVD Scores for Seasonal Spatial Variance Maps}} \\
    \midrule
    & \textbf{2 steps} & \textbf{4 steps} & \textbf{6 steps} & \textbf{8 steps} & \textbf{10 steps} & \textbf{12 steps} & \textbf{14 steps} \\
    \cmidrule{2-8}
    \textbf{yearly} & $1.88\text{e-01}$ & $1.45\text{e-01}$ & $1.41\text{e-01}$ & $\underline{1.40\text{e-01}}$ & $1.43\text{e-01}$ & $1.45\text{e-01}$ & $1.49\text{e-01}$ \\
    \midrule
    \textbf{JFM} & $1.14\text{e-01}$ & $\underline{1.02\text{e-01}}$ & $1.04\text{e-01}$ & $1.10\text{e-01}$ & $1.19\text{e-01}$ & $1.27\text{e-01}$ & $1.32\text{e-01}$ \\
    \textbf{AMJ} & $2.15\text{e-01}$ & $1.62\text{e-01}$ & $1.42\text{e-01}$ & $1.37\text{e-01}$ & $1.34\text{e-01}$ & $\underline{1.33\text{e-01}}$ & $1.34\text{e-01}$ \\
    \textbf{JAS} & $2.22\text{e-01}$ & $1.74\text{e-01}$ & $1.58\text{e-01}$ & $1.50\text{e-01}$ & $1.47\text{e-01}$ & $\underline{1.46\text{e-01}}$ & $1.47\text{e-01}$ \\
    \textbf{OND} & $1.42\text{e-01}$ & $\underline{1.10\text{e-01}}$ & $1.11\text{e-01}$ & $1.17\text{e-01}$ & $1.24\text{e-01}$ & $1.29\text{e-01}$ & $1.35\text{e-01}$ \\
    \bottomrule
\end{tabular}
\captionsetup{width=0.94\textwidth}
\caption{Mean variance discrepancy (MVD) scores computed between the spatial mean variance maps of Ensemble Diffusion and CERRA-EDA for each three-monthly period (JFM, AMJ, JAS, OND) and each timestamp for 2016. Lower scores indicate higher similarity between the variance distributions of the two ensembles.}
\label{tab:MVD_between_variance_images}
\end{table*}

\begin{figure*}[ht]
    \centering
    \includegraphics[width=\textwidth]{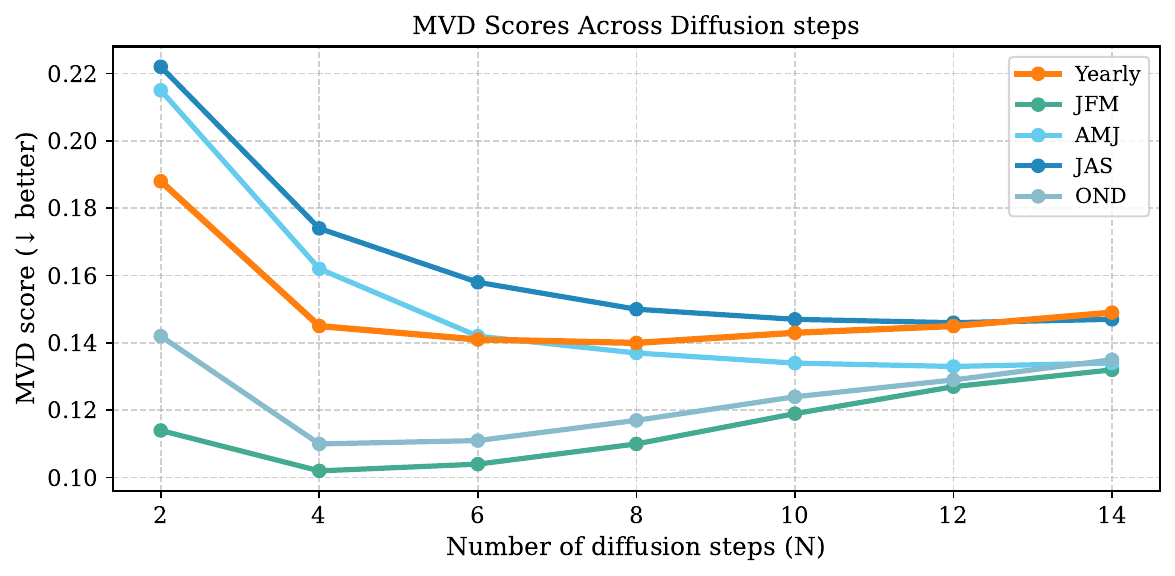}
    \caption{Evolution of the mean variance discrepancy (MVD) scores with the number of diffusion steps. The plot visualizes the values reported in Table \ref{tab:MVD_between_variance_images} for the yearly and three-monthly periods (JFM, AMJ, JAS, OND) during 2016. Lower values indicate a closer match between the spatial variance distributions of Ensemble Diffusion and CERRA-EDA. The yearly curve (orange) highlights the global trend, while the seasonal curves illustrate the differing optimal step counts across the annual cycle.}
    \label{fig:MVD_scores}
\end{figure*}

\subsection{Impact of \texorpdfstring{$\Delta t$}{Delta t} on performance}
An interesting aspect of the variance analyses is to check how the overall performance of the Ensemble Diffusion (ED) model changes as the number of steps increase. To achieve this goal we have have utilized $M=10$ ensemble members at $565 \times 565$ spatial resolution and 6-hourly time resolution for the year 2016. We have computed the ensemble diffusion mean and compared it with the available CERRA-EDA control run, which is, as mentioned in Section \ref{subsec:CERRA_EDA}, the member of the ensemble with no perturbation on the model input. We evaluate the comparison via MSE, calculated on normalized values. The results, reported in table \ref{tab:performance_change_timestamp}, reveal that the performance is largely stable while changing the step size $\Delta t$, with a slight increase in the performance for larger step sizes. This should be expected as more steps allows the reverse diffusion process to make smaller, more gradual corrections at each step, leading to a more precise reconstruction of the target high-resolution image.

\begin{figure*}[ht]
    \centering
    \includegraphics[width=\textwidth]{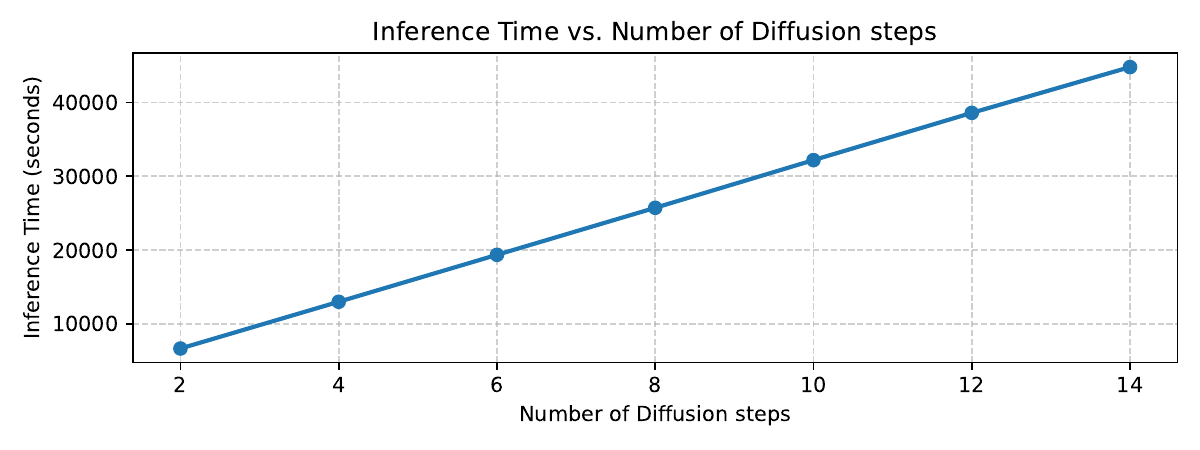}
    \caption{Inference time of the ensemble diffusion model as a function of the number of diffusion steps. The results show an approximately linear relationship, confirming that computational cost scales proportionally with the number of reverse diffusion iterations.}
    \label{fig:inference_times}
\end{figure*}

For what concerns execution efficiency, the number of diffusion steps is expected to scale linearly with the total inference time. Under an inference setup analogous to the training configuration described in \Cref{subsec:fulldomain_training} (three NVIDIA A100 GPUs) the time required to generate 10 ensemble members for the year 2016 at a 6-hour temporal resolution ranges from approximately 110 minutes for 2 diffusion steps to 746 minutes for 14 steps. The relationship between inference time and the number of diffusion steps is illustrated in \Cref{fig:inference_times}.

\begin{table*}[htbp]
\centering
\def\arraystretch{1.1} % for increased vertical spacing
\small % or \footnotesize if still too wide
\begin{tabular}{>{\bfseries}lcccccccc}
    \toprule
    \multicolumn{8}{c}{\textbf{MSE ensemble mean performance comparison}} \\
    \midrule
    & \textbf{2 steps} & \textbf{4 steps} & \textbf{6 steps} & \textbf{8 steps} & \textbf{10 steps} & \textbf{12 steps} & \textbf{14 steps} \\
    \cmidrule{2-8}
    \textbf{yearly} & $1.67\text{e-03}$ & $1.62\text{e-03}$ & $1.60\text{e-03}$ & $1.59\text{e-03}$ & $1.58\text{e-03}$ & $1.58\text{e-03}$ & $\underline{1.57\text{e-03}}$ \\
    \midrule
    \textbf{JFM} & $1.98\text{e-03}$ & $1.92\text{e-03}$ & $1.90\text{e-03}$ & $1.88\text{e-03}$ & $1.87\text{e-03}$ & $1.86\text{e-03}$ & $\underline{1.86\text{e-03}}$ \\
    \textbf{AMJ} & $1.52\text{e-03}$ & $1.48\text{e-03}$ & $1.47\text{e-03}$ & $1.46\text{e-03}$ & $1.45\text{e-03}$ & $1.45\text{e-03}$ & $\underline{1.44\text{e-03}}$ \\
    \textbf{JAS} & $1.39\text{e-03}$ & $1.36\text{e-03}$ & $1.34\text{e-03}$ & $1.33\text{e-03}$ & $1.33\text{e-03}$ & $1.33\text{e-03}$ & $\underline{1.32\text{e-03}}$ \\
    \textbf{OND} & $1.79\text{e-03}$ & $1.73\text{e-03}$ & $1.70\text{e-03}$ & $1.68\text{e-03}$ & $1.68\text{e-03}$ & $\underline{1.67\text{e-03}}$ & $\underline{1.67\text{e-03}}$ \\
    \bottomrule
\end{tabular}
\captionsetup{width=0.94\textwidth}
\caption{Table comparing the MSE performance of the diffusion model at different numbers of reverse diffusion steps. The MSE is calculated between the mean of the 10 generated ensembles and the CERRA-EDA control run. The results are reported 3-monthly for the year 2016, along with the yearly mean across all seasons.}
\label{tab:performance_change_timestamp}
\end{table*}

\begin{figure*}[ht]
    \centering
    \includegraphics[width=\textwidth]{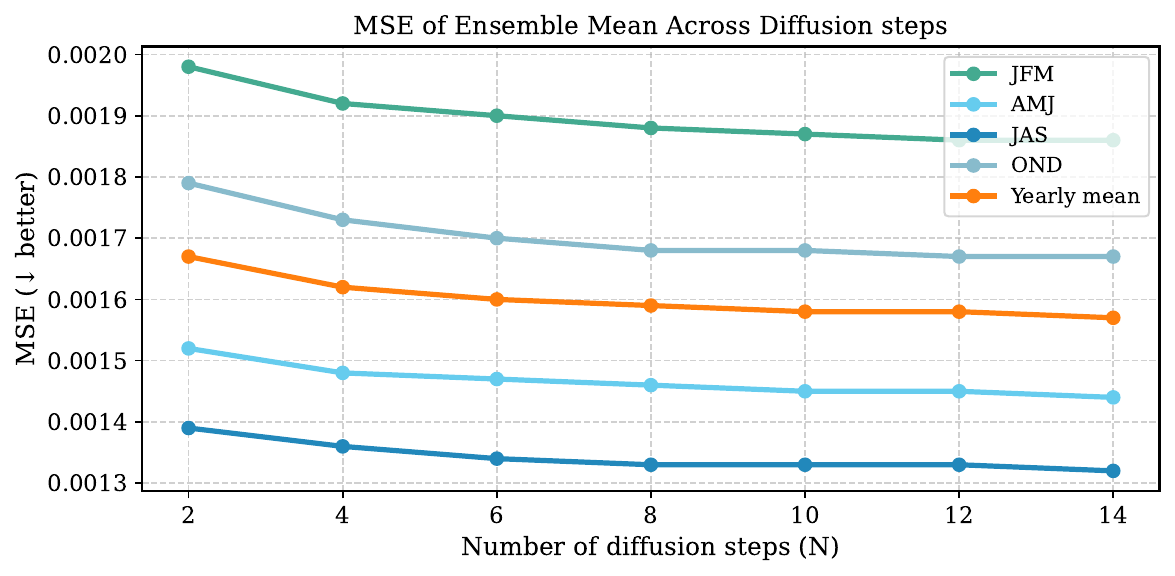}
    \caption{Mean squared error (MSE) of the ensemble mean versus the CERRA-EDA control run for different diffusion steps. Results for each three-monthly period (JFM, AMJ, JAS, OND) in 2016 show stable performance with slightly lower errors at higher step counts.}
    \label{fig:MSE_across_steps}
\end{figure*}

% \subsection{variance in a single point}
% Alongside full domain analyses it could be significative to analyze the behaviour of the generate ensamble for a single point. For our analyses we chose a random point of the map and plot the mean value alongside a variance range, for both ensemble diffusion with 8 steps and CERRA-EDA. To achieve a better clarity with a lenghty time domain we applied a moving average smoothing with a window of 30 samples. The resulting image is reported in Figure \ref{fig:single_point_variance}. The plot reveal consistent overlap between mean and variance ranges.

\subsection{Demonstrating capabilities on CARRA}
Both CERRA and CARRA, introduced in Section \ref{sec:Data} are built upon the HARMONIE-AROME weather prediction system \cite{bengtsson2017harmonie}, adapted for reanalysis purposes. This common foundation ensures that both systems share a core methodology for atmospheric modeling and data assimilation, making them methodologically similar despite their focus on different regions.

Having validated our diffusion model's capability to generate ensemble members for the CERRA dataset, we naturally extend this methodology to CARRA-EAST, the Arctic region dataset focusing on the eastern domain. Unlike CERRA, CARRA lacks an ensemble dataset, which limits its usability in quantifying variability, validating against real-world uncertainties, detecting rare or extreme events, performing ensemble-based statistical analyses, and estimating model robustness.
Our ensemble diffusion driven ensemble dataset can be a highly efficient way to address this issues. 

For CARRA-EAST, we adopt a training setup similar to the one proposed for CERRA. The downscaling process is framed as an image-to-image transformation, converting low-resolution ERA5 data into high-resolution CARRA-EAST, mantaining the same spatial domain between the two, with the resulting grid having a size of  $989\times 789$. We regrid ERA5 wind speed data to match the projection of CARRA-EAST. The training period spans 1991–2010, while the year 2016 is reserved for ensemble testing. As with CERRA, padding is applied to adjust the grid size for compatibility with the diffusion architecture. Training was conducted following the setup detailed in Section \ref{subsec:fulldomain_training}.

\begin{figure*}
    \centering
    \includegraphics[width=\textwidth]{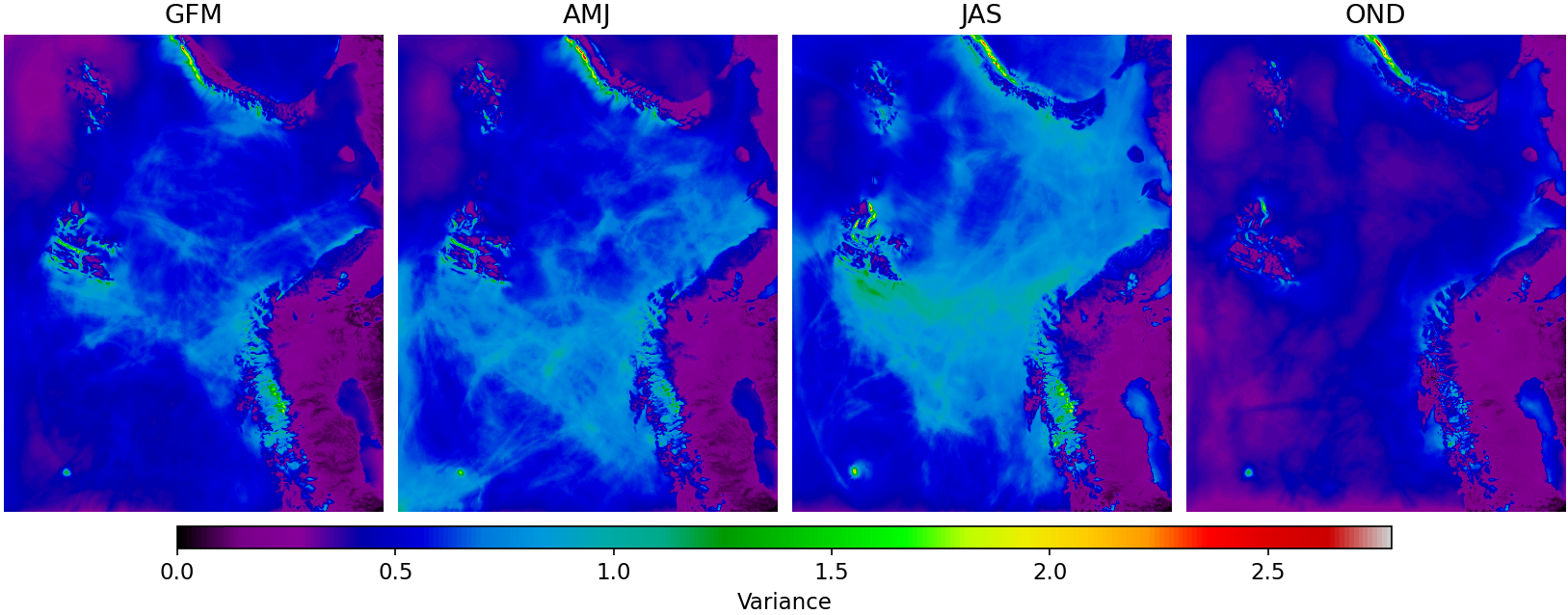}
    \caption{Revealing the spatial variance of ensemble diffusion trained on CARRA-EAST with 8 diffusion steps. The comparison is relative to the testing year 2016 and is performed in a 3 monthly manner to highlight seasonal variations.}
    \label{fig:carra_var}
\end{figure*}

The ensemble members are generated for the year 2016 using 8 diffusion steps, the optimal value determined through spatial validation with MVD on CERRA-EDA. Figure \ref{fig:carra_var} illustrates the spatial variance for 2016, plotted within the same variance range as Figure \ref{fig:spatial_variance_big_picture}. Our results reveal that the spatial variance is consistent with expected patterns. 

\subsection{Comparison with EMOS}

Our method provides a direct ensemble comparison with CERRA-EDA, the numerically computed ensemble of CERRA, which can be naturally regarded as the optimal ground truth for evaluation. To ensure a comprehensive evaluation, our method can be positioned alongside other ensemble-generation approaches, including well-established statistical post-processing techniques such as EMOS.
Among neural stochastic architectures, methods based on GANs have been proposed for related tasks, yet their training and tuning requirements make them impractical in our setting, as discussed in \Cref{subsec:reanalysis_downscaling_DL}. For this reason, our comparison focuses on EMOS.

EMOS (Ensemble Model Output Statistics) is a statistical post-processing approach widely used to calibrate ensemble and deterministic forecasts \cite{gneiting2005calibrated}. Originally developed as a calibration framework for correcting bias and spread in ensemble systems, EMOS has become one of the most adopted post-processing techniques in numerical weather prediction and downscaling contexts \cite{thorarinsdottir2010probabilistic, baran2016mixture, szabo2023parametric}. Beyond calibration, EMOS can also be applied to generate probabilistic ensembles directly from deterministic forecasts by modeling the predictive distribution of the target variable as a function of the forecast itself. 
This formulation allows EMOS to correct systematic biases and rescale forecast variability, producing calibrated ensemble predictions that are statistically consistent with the observed data.

In our setting, EMOS is applied to the deterministic outputs of the reference U-Net to generate probabilistic ensembles. We employ a deterministic version of our downscaling model, a high-performing residual U-Net, which achieves performance comparable to the diffusion model in terms of MSE for the testing years, as reported in \Cref{tab:metrics_comparison}. 
This deterministic U-Net provides an ideal baseline for comparison, and our experiment focuses on generating ensemble members from the U-Net using EMOS and evaluating them against those obtained from the diffusion-based ensembles.

The objective of this experiment is to evaluate whether EMOS, when fitted to the relationship between the deterministic forecasts and the ground truth, can reproduce the ensemble variability observed in CERRA-EDA. 
In practice, EMOS models the conditional distribution of wind speed as the truncated normal distribution,
$Y \sim \mathcal{TN}_0(\mu = a + bX, \sigma)$,
where the coefficients $a$ and $b$ correct systematic bias and amplitude scaling, and $\sigma$ controls the ensemble spread. 
The deterministic U-Net, trained on the 1985–2010 period (consistent with the diffusion model), was used as the forecast source. 
The year 2015 was used to fit the EMOS parameters against the CERRA ground truth, while 2016 was reserved for evaluation against the ten-member CERRA-EDA ensemble. 
All datasets were normalized to the $[0,1]$ range using the CERRA training statistics. 
EMOS fitting was performed after bilinearly downscaling the raw U-Net outputs to $565 \times 565$ pixels to match the spatial resolution of CERRA-EDA. 
The evaluation was carried out at a $6$-hourly frequency and $565 \times 565$ spatial resolution, comparing the EMOS-generated ensemble with the reference ensemble dataset.

\begin{table}[htbp]
\centering
\def\arraystretch{1.1}
\begin{tabular}{>{\bfseries}lc}
    \toprule
    \multicolumn{2}{c}{\textbf{Total Ensemble Variance Comparison}} \\
    \midrule
    \textbf{2 steps}       & $0.216$ \\
    \textbf{4 steps}       & $0.298$ \\
    \textbf{6 steps}       & $0.338$ \\
    \textbf{8 steps}       & $0.357$ \\
    \textbf{10 steps}      & $0.375$ \\
    \textbf{12 steps}      & $0.388$ \\
    \textbf{14 steps}      & $0.397$ \\
    \midrule
    \textbf{EMOS}          & $0.392$ \\
    \textbf{Local EMOS}    & $0.392$ \\
    \textbf{CERRA-EDA (Ref.)} & $\mathbf{0.392}$ \\
    \bottomrule
\end{tabular}
\captionsetup{width=0.94\textwidth}
\caption{Total ensemble variance ($\mu_V$) for Ensemble Diffusion at increasing numbers of reverse diffusion steps, compared with EMOS, Local EMOS, and the reference CERRA-EDA variance. Variance increases monotonically with diffusion steps and converges toward the reference.}
\label{tab:total_variance_comparison}
\end{table}

Two EMOS configurations were tested. 
In the standard EMOS, a single set of parameters $(a,b,\sigma)$ was fitted over the entire 2015 dataset, yielding $a=-7.1\times10^{-4}$, $b=1.015$, and $\sigma=0.0168$. 
The spread was subsequently calibrated to match the global mean variance of CERRA-EDA via  
$\sigma_{\mathrm{new}} = \sigma \sqrt{\mu V_{\mathrm{target}} / \mu V_{\mathrm{current}}}$,  
resulting in $\sigma_{\mathrm{new}}=0.0122$. 
In the local EMOS, a separate regression was fitted for each grid point,
$Y_t(i,j)=a_{i,j}+b_{i,j}X_t(i,j)+\varepsilon_t(i,j)$,
producing spatial maps of $a_{i,j}$, $b_{i,j}$, and $\sigma_{i,j}$. 
To prevent over-dispersion, the per-pixel spread was globally scaled using the same variance calibration factor,  
$k=\sqrt{\mu V_{\mathrm{target}}/\mu V_{\mathrm{current}}}\approx0.60$.

\begin{table}[htbp]
\centering
\def\arraystretch{1.1} % for increased vertical spacing
\begin{tabular}{>{\bfseries}lccc}
    \toprule
    \multicolumn{4}{c}{\textbf{MVD Scores for Ensemble Diffusion (8 steps) vs EMOS}} \\
    \midrule
    & \textbf{8 steps (Ours)} & \textbf{EMOS} & \textbf{Local EMOS} \\
    \cmidrule{2-4}
    \textbf{Yearly} & $\underline{1.40\text{e-01}}$ & $1.60\text{e-01}$ & $1.59\text{e-01}$ \\
    \midrule
    \textbf{JFM} & $\underline{1.11\text{e-01}}$ & $1.65\text{e-01}$ & $1.59\text{e-01}$ \\
    \textbf{AMJ} & $\underline{1.37\text{e-01}}$ & $1.51\text{e-01}$ & $1.52\text{e-01}$ \\
    \textbf{JAS} & $1.51\text{e-01}$ & $\underline{1.40\text{e-01}}$ & $1.44\text{e-01}$ \\
    \textbf{OND} & $\underline{1.17\text{e-01}}$ & $1.43\text{e-01}$ & $1.40\text{e-01}$ \\
    \bottomrule
\end{tabular}
\captionsetup{width=0.94\textwidth}
\caption{Mean variance discrepancy (MVD) scores computed between the spatial mean variance maps of the 8-step Ensemble Diffusion model, EMOS, and Local EMOS for each three-monthly period (JFM, AMJ, JAS, OND) and the full year 2016. Lower scores indicate higher similarity between the variance distributions of the compared ensembles.}
\label{tab:MVD_emos_comparison}
\end{table}

Naturally, both EMOS variants successfully reproduce the total ensemble variance of CERRA-EDA, as they are directly calibrated to match this value, as reported in \Cref{tab:total_variance_comparison}. 
The more interesting question is whether they can approximate the \emph{spatial distribution} of the variance. 
For the standard EMOS approach, which relies on a single value of $\sigma$, this is inherently impossible. 
However, the local EMOS formulation, which estimates parameters for each grid point, may be able to capture some spatial structures. 
Having computed spatial variance maps analogous to those obtained for the diffusion ensembles, the local EMOS results were added to \Cref{fig:spatial_variance_big_picture}. 
As shown in the figure, EMOS identifies a few regions of high variability, but these are limited in extent and fail to represent the complex spatial patterns of variance clearly captured by both CERRA-EDA and the diffusion model.

The mean-variance discrepancy (MVD) scores computed between EMOS, local EMOS, and the best-performing diffusion configuration (8 steps), reported in \Cref{tab:MVD_emos_comparison}, confirm this observation. 
In the yearly comparison, the diffusion model clearly outperforms both EMOS configurations, and the same holds for three of the four seasonal partitions. 
Interestingly, however, during the July–August–September period, the EMOS configuration slightly outperforms diffusion. 
This finding is noteworthy, given that EMOS is a purely statistical method that does not explicitly model spatial variability. 
This outcome likely reflects a relative reduction in the diffusion model’s ability to capture uncertainty during the summer months, rather than a genuine advantage of EMOS, suggesting that ensemble diffusion may struggle to represent variance under the distinct dynamical conditions of this season.

\subsection{Limitations}

While the proposed framework demonstrates that ensemble diffusion models can effectively reproduce the statistical variability of regional wind fields, several aspects merit further exploration.
First, the analysis has been limited to wind speed. Although preliminary tests on other variables such as temperature and precipitation yielded comparable results, suggesting that the core principles of our approach (variance calibration and spatial consistency evaluation) may generalize beyond wind-related fields, a comprehensive multi-variable analysis would provide valuable insight into the universality and robustness of the proposed methodology, including any variable-specific adjustments that may be required.
The validation of ensemble variance relied on CERRA-EDA as the reference dataset. While this ensemble offers a consistent and physically grounded benchmark, it operates at a reduced spatial and temporal resolution and thus may not capture the full spectrum of uncertainty present in observations. Future studies could strengthen the evaluation by incorporating in-situ measurements, alternative reanalysis ensembles, or observational uncertainty estimates.
Geographically, the experiments focused on the European CERRA domain with an extension to CARRA-EAST. Although these regions encompass diverse climatological regimes, confirming the model’s robustness across other continents and observational contexts (particularly areas with different dynamics or sparser data) would further consolidate the method’s general applicability.
Finally, the evaluation primarily targeted variance alignment through global metrics such as mean variance deviation (MVD). While this provides a first-order measure of ensemble consistency, complementary diagnostics, including spread–skill relationships, reliability diagrams, and extreme-event statistics, could offer a more comprehensive assessment of ensemble quality.
Overall, these limitations delineate promising directions for future work. Extending the framework to additional variables, reference datasets, and regions, as well as enriching the evaluation protocols, will further clarify the potential of diffusion-based ensembles as physically interpretable and controllable tools for climate downscaling.

\section{Conclusions}\label{sec:Conclusion}
In this work, we applied an ensemble diffusion model to the full-domain ERA5-to-CERRA downscaling task and discovered a relationship between the variance exhibited by the ensemble diffusion and the number of reverse diffusion steps, giving theoretical proof of this relation. Notably, we demonstrated that the selection of the reverse diffusion step size $\Delta t$ not only impacts the time efficiency of the generative process but also serves as a crucial mechanism for controlling the variance of the sampled data, a key aspect when using these models for generating ensembles. We have explored architectural improvements needed to apply ensemble diffusion models over large domains, including the use of a scaling factor to reduce the signal rate in the noise scheduler, reducing the effect that higher resolution have on denoising difficulty. Our experimental results indicate that adjusting the number of diffusion steps enables the model to operate across different ranges of ensemble variance. We found that the control over variance gained by varying the number of diffusion steps is sufficient to match the global mean variance exhibited by the available ensemble reference dataset, CERRA-EDA. We observed that increasing the number of diffusion steps leads to convergence toward a fixed variance value. Although this converged global mean variance closely matches the one exhibited by CERRA-EDA, our MVD score, which takes into account the spatial distribution of the variance, revealed that the optimal number of steps is 8, a value far from the convergence point. This demonstrates that selecting a ``sufficiently high'' number of steps is not necessarily an optimal strategy for tuning an ensemble diffusion model, highlighting the importance of calibration to obtain optimal results. 

To further contextualize the effectiveness of our approach, we have also introduced an Ensemble Model Output Statistics (EMOS) baseline as a comparison method. EMOS, a widely adopted statistical post-processing technique for ensemble calibration, provided a valuable reference for assessing the seasonal variability and overall performance of our diffusion model. While EMOS successfully reproduced global variance levels, it lacked the spatial variance coherence achieved by the diffusion-based approach, underscoring the advantages of learning-based generative methods for ensemble generation.

In our work we have showcased the capability of generating ensemble members efficiently at high temporal and spatial resolutions. Specifically, our ensemble reference CERRA-EDA has a considerable reduction in spatial resolution (from $1069 \times 1069$ to $565 \times 565$) and temporal resolution (from $3h$ to $6h$). We reveal that our approach is able to generate ensemble members at full CERRA temporal and spatial resolution, while maintaining high efficiency, essentially running on workstation hardware. Considering that our model is conditioned only by ERA5 low-resolution inputs, which happens to be available at hourly intervals, we can additionally increase the ensemble temporal resolution to $1h$. Additionally, we have extended our ensemble diffusion methodology from CERRA to CARRA-EAST, demonstrating its capability to generate ensemble members for datasets lacking ensemble information, using a similar training setup and confirming its effectiveness in reproducing spatial variance patterns through diffusion step tuning.

Since our diffusion model is trained solely on the deterministic ERA5-to-CERRA task, it is unaffected by factors such as measurement quality or other previously assessed uncertainties. As such, our approach can serve as a validation tool for existing ensembles, with discrepancies providing insights into the neural model's understanding of the underlying physical system or serving as validation for uncertainty estimates obtained through other methods. 

In our experimental setting, the spatial variance reported in Figure \ref{fig:spatial_variance_big_picture} shows a clear overlap between distributions. However, during the summer months in the North African region, discrepancies between CERRA-EDA and our Ensemble Diffusion appear. This may stem from physical factors limiting the model’s ability to learn regional variance dynamics, particularly under intense surface heating, shallow convective mixing, and heterogeneous land-surface conditions typical of the summer season. Alternatively, it may reflect inflated variance in CERRA-EDA itself, driven by data scarcity and uncertainty inflation in areas with sparse observations. Although a full investigation of these mechanisms lies beyond the scope of this work, we note that a season-specific tuning of the number of reverse diffusion steps mitigates this issue.

In conclusion, our methodology provides a practical tool for tuning ensemble diffusion models to produce the correct variance distribution. We believe that future implementations should consider the number of diffusion steps as a crucial tuning parameter. Furthermore, our theoretical framework opens promising directions for studying the statistical properties of ensemble members in relation to the structure of diffusion models, encouraging further research.

\section{Acknowledgements}
ERA5, CERRA, CERRA-EDA and CARRA were downloaded from the Copernicus Climate Change Service (C3S) (2023). This research was partially funded and supported by the following Projects:
\begin{itemize}
\item European Cordis Project ``Optimal High Resolution Earth System Models for Exploring Future Climate Changes'' (OptimESM),
Grant agreement ID: 101081193
% \item Future AI Research (FAIR) project of the National Recovery and Resilience Plan (NRRP), Mission 4 Component 2 Investment 1.3 funded from the European Union - NextGenerationEU.
\item ISCRA Project ``AI for weather analysis and forecast'' (AIWAF)
\end{itemize}

\subsection*{Statements and Declarations}
The authors declare no competing interests.

\subsection*{Code availability}
Code availability removed to ensure blindness.
\bibliography{bibliography}

\appendix
\section{Proofs of main results}\label{app:proofs}
In this section, we report the derivation of the main results presented in the paper.

\subsection{Full proof of Theorem~\ref{teo:closed_formula} (scalar case)}\label{app:scalar_proof}
For completeness, we report here the full derivation of Theorem~\ref{teo:closed_formula}, 
whose main steps were only sketched in the paper body. 
The proof below includes all intermediate Taylor expansions and variance–recursion arguments.

\begin{proof}
    Let $\alpha(t)$ be the continuous extension of $\{ \alpha_t\}_{t = 1}^T$. 
    The first-order Taylor decomposition of $\sqrt{\alpha(t)}$ around $t-\Delta t$ shows:
    \begin{align*}
        \sqrt{\alpha(t)} 
        &= \sqrt{\alpha(t-\Delta t)} 
        -  \left(\sqrt{\alpha(t-\Delta t)}  \right)' \Delta t 
        + O(\Delta t^2) \\[1mm]
        &= \sqrt{\alpha(t-\Delta t)} 
        - \frac{\alpha'(t - \Delta t)}{2\sqrt{\alpha(t - \Delta t)}}\,\Delta t 
        + O(\Delta t^2),
    \end{align*}
    and similarly, for $\sqrt{1 - \alpha(t)}$:
    \begin{align*}
        \sqrt{1 - \alpha(t)} 
        &= \sqrt{1 - \alpha(t-\Delta t)} 
        -  \left(\sqrt{1 - \alpha(t-\Delta t)}  \right)' \Delta t 
        + O(\Delta t^2) \\[1mm]
        &= \sqrt{1 - \alpha(t-\Delta t)} 
        + \frac{\alpha'(t - \Delta t)}{2\sqrt{1 - \alpha(t - \Delta t)}}\,\Delta t 
        + O(\Delta t^2).
    \end{align*}
    Consequently:
    \begin{align*}
        \sqrt{\frac{\alpha(t)}{\alpha(t-\Delta t)}} 
        = 1 - \frac{\alpha'(t - \Delta t)}{2 \alpha(t-\Delta t)}\, \Delta t 
        + O(\Delta t^2),
    \end{align*}
    and
    \begin{align*}
        c_{t - \Delta t} 
        &= \sqrt{1 - \alpha(t)} 
        - \sqrt{\alpha(t)}\sqrt{\frac{(1 - \alpha(t-\Delta t))}{\alpha(t-\Delta t)}} \\[1mm]
        &= \sqrt{1 - \alpha(t)} 
        - \sqrt{1 - \alpha(t-\Delta t)} 
        + \alpha'(t - \Delta t)
          \frac{\sqrt{1 - \alpha(t - \Delta t)}}{2\alpha(t - \Delta t)}\, \Delta t 
        + O(\Delta t^2) \\[1mm]
        &= \frac{\alpha'(t - \Delta t)}{2\sqrt{1 - \alpha(t - \Delta t)}}\, \Delta t 
        + \alpha'(t - \Delta t)
          \frac{\sqrt{1 - \alpha(t - \Delta t)}}{2\alpha(t - \Delta t)}\, \Delta t 
        + O(\Delta t^2) \\[1mm]
        &= \left( 
            \frac{1}{2\sqrt{1 - \alpha(t - \Delta t)}} 
            + \frac{\sqrt{1 - \alpha(t - \Delta t)}}{2\alpha(t - \Delta t)}
           \right) 
           \alpha'(t - \Delta t)\, \Delta t 
           + O(\Delta t^2).
    \end{align*}

    Plugging these quantities into the definition of 
    $\boldsymbol{F}_t$ and $\boldsymbol{g}_t$ 
    as in Proposition \ref{prop:recursive_formula} yields:
    \begin{align*}
        \boldsymbol{F}_t 
        &= \left( 
            \sqrt{\frac{\alpha(t)}{\alpha(t - \Delta t)}} 
            + c_{t - \Delta t} \boldsymbol{J}_{t - \Delta t} 
           \right)^2 \\[2mm]
        &= \Bigg[ 
            1 
            - \Bigg(
              \frac{1}{2 \alpha(t-\Delta t)} 
              - \frac{\sqrt{1 - \alpha(t - \Delta t)}}{2 \alpha(t-\Delta t)} 
                \boldsymbol{J}_{t - \Delta t} 
              + \frac{1}{2\sqrt{1 - \alpha(t - \Delta t)}}
                \boldsymbol{J}_{t - \Delta t}
            \Bigg) \\[1mm]
        &\hspace{220px} \cdot \alpha'(t - \Delta t)\, \Delta t 
            + O(\Delta t^2)
          \Bigg]^2 \\[2mm]
        &= 1 
        + \Bigg[
            - \frac{\alpha'(t - \Delta t)}{\alpha (t - \Delta t)} 
            + \boldsymbol{J}_{t - \Delta t}\, \alpha'(t - \Delta t)
              \frac{\sqrt{1 - \alpha(t - \Delta t)}}{\alpha(t - \Delta t)} \\[1mm]
        &\hspace{180px} + \boldsymbol{J}_{t - \Delta t}\, 
              \frac{\alpha'(t - \Delta t)}{\sqrt{1 - \alpha(t - \Delta t)}}
          \Bigg] \Delta t 
          + O(\Delta t^2) \\[1mm]
        &= 1 
          +
            \Bigg(
              - \partial_t \log \alpha(t - \Delta t) 
              + \boldsymbol{J}_{t - \Delta t}\, \alpha'(t - \Delta t)
                \frac{\sqrt{1 - \alpha(t - \Delta t)}}{\alpha(t - \Delta t)} \\[1mm]
        &\hspace{170px} + \boldsymbol{J}_{t - \Delta t}\,
                \frac{\alpha'(t - \Delta t)}{\sqrt{1 - \alpha(t - \Delta t)}}
            \Bigg)
          \Delta t 
          + O(\Delta t^2) \\[2mm]
        &= 1 + \lambda_{t - \Delta t} \Delta t + O(\Delta t^2).
    \end{align*}
    and
    \begin{align*}
        \boldsymbol{g}_t 
        &= c_{t - \Delta t}^2 \ewVar(\boldsymbol{r}_t) \\[1mm]
        &=  \underbrace{
             \left(
               \frac{\alpha'(t - \Delta t)\sqrt{1 - \alpha(t - \Delta t)}}{2\alpha(t - \Delta t)} 
               +\frac{\alpha'(t - \Delta t)}{2\sqrt{1 - \alpha(t - \Delta t)}}
             \right) 
             \TotVar(\boldsymbol{r}_t)
           }_{:=\gamma_{t - \Delta t}}
           \Delta t 
           + O(\Delta t^2) \\[1mm]
        &= \gamma_{t- \Delta t} \Delta t + O(\Delta t^2).
    \end{align*}
    To show monotonicity of the sequence 
    $\{ \TotVar_{N \Delta t} \}_{N \in \mathbb{N}}$, 
    note that the above derivation, together with the observation 
    that $\Delta t \to 0$ as $N \to \infty$, implies that:
    \begin{align}\label{eq:monotonic_product}
        \begin{aligned}
            \log \prod_{i = 1}^N \boldsymbol{F}_{i\Delta t} 
            &= \sum_{i=1}^N 
               \log \left( 1 + \lambda_{t - \Delta t}\, \Delta t + O(\Delta t^2) \right) \\[1mm]
            &\approx \sum_{i=1}^N 
               \lambda_{t - \Delta t}\, \Delta t + O(\Delta t^2)
               \overset{\Delta t \to 0}{\longrightarrow} 
               \int_{0}^\infty \lambda(t)\,dt \\[1mm]
            \iff 
            \prod_{i = 1}^N \boldsymbol{F}_{i\Delta t} 
            &\overset{\Delta t \to 0}{\longrightarrow} 
             e^{\int_{0}^\infty \lambda(t)\,dt},
        \end{aligned}
    \end{align}
    where the convergence is monotonic from below due to 
    standard Euler summation arguments. 
    Therefore, by Equation \eqref{eq:recursive_formula}, 
    the $N$-th term of the sequence 
    $\{ \TotVar_{N \Delta t} \}_{N \in \mathbb{N}}$ reads:
    \begin{align*}
        \v_{N \Delta t} 
        &\approx 
          \left(\prod_{i = 1}^N \boldsymbol{F}_{i \Delta t} \right)
          + \sum_{i = 1}^N 
            \left[
              \left( \prod_{k = i+1}^{N+1} 
                     \boldsymbol{F}_{k \Delta t} \right) 
              \boldsymbol{g}_{i \Delta t}
            \right],
    \end{align*}
    and consequently:
    \begin{align*}
        \v_{(N+1) \Delta t} 
        &\approx 
          \left(\prod_{i = 1}^{N+1} \boldsymbol{F}_{i \Delta t} \right) 
          + \sum_{i = 1}^{N+1} 
            \left[
              \left( \prod_{k = i+1}^{N+2} 
                     \boldsymbol{F}_{k \Delta t} \right) 
              \boldsymbol{g}_{i \Delta t}
            \right] \\[1mm]
        &\geq 
          \left(\prod_{i = 1}^N \boldsymbol{F}_{i \Delta t} \right) 
          + \sum_{i = 1}^{N+1} 
            \left[
              \left( \prod_{k = i+1}^{N+1} 
                     \boldsymbol{F}_{k \Delta t} \right) 
              \boldsymbol{g}_{i \Delta t}
            \right]
          \quad \text{(by Equation \eqref{eq:monotonic_product})} \\[1mm]
        &= \left(\prod_{i = 1}^N \boldsymbol{F}_{i \Delta t} \right) 
           + \sum_{i = 1}^N 
             \left[
               \left( \prod_{k = i+1}^{N+1} 
                      \boldsymbol{F}_{k \Delta t} \right) 
               \boldsymbol{g}_{i \Delta t}
             \right] 
           + \boldsymbol{g}_{(N+1) \Delta t} \\[1mm]
        &= \TotVar_{N\Delta t} + \boldsymbol{g}_{(N+1) \Delta t} \\[1mm]
        &\geq \TotVar_{N\Delta t}
           \quad \text{(since } \boldsymbol{g}_{(N+1) \Delta t} \geq 0\text{)},
    \end{align*}
    proving monotonicity. 
    Arguments similar to \eqref{eq:monotonic_product} 
    also show that:
    \begin{align*}
        \TotVar_{N\Delta t} 
        \overset{\Delta t \to 0}{\longrightarrow} 
        e^{- \int_0^T \lambda(s)\, ds} 
        + \int_0^T 
            e^{- \int_0^s \lambda(\kappa)\,d\kappa} 
            \gamma(s)\, ds,
    \end{align*}
    concluding the proof.
\end{proof}

\section{Extension to the \texorpdfstring{$n$}{n}-Dimensional Case}\label{app:ndim_proof}

In the main text, we presented the complete theoretical analysis for the scalar case ($n=1$), which already provides an intuitive understanding of the evolution of the variance across DDIM timesteps. 
For completeness, in this Appendix we report the extended derivation for the general $n$-dimensional case, where $\x_t \in \mathbb{R}^n$. 
The same reasoning applies, but with the variance replaced by the full covariance matrix, and the Jacobian $\boldsymbol{J}_t$ promoted to a matrix. 
We then show that the total variance, defined as the trace of the covariance matrix, follows an analogous affine recursion and converges monotonically to an exponential-affine limit.

\begin{proposition}\label{prop:recursive_formula_ndim}
    Let $\boldsymbol{m}_t:= \mathbb{E}[\x_t]$ and $\boldsymbol{\Sigma}_t := Var(\x_t)$. 
    Then, under the first-order linearization of the DDIM update rule, it holds:
    \begin{align}\label{eq:recursive_formula_ndim}
        \boldsymbol{\Sigma}_t \approx \boldsymbol{A}_t \boldsymbol{\Sigma}_{t - \Delta t} \boldsymbol{A}_t^T + c_{t - \Delta t}^2 Var(\boldsymbol{r}_t),
    \end{align}
    where:
    \begin{align*}
        \boldsymbol{A}_t := \sqrt{\frac{\alpha_t}{\alpha_{t - \Delta t}}}\,\boldsymbol{I} + c_{t - \Delta t}\boldsymbol{J}_{t - \Delta t},
    \end{align*}
    and $\boldsymbol{r}_t$ is a random vector independent of $\x_{t-\Delta t}$.
\end{proposition}

\begin{proof}
    Starting from the DDIM step
    \begin{align*}
        \x_t = \sqrt{\frac{\alpha_t}{\alpha_{t - \Delta t}}} \x_{t - \Delta t} + c_{t - \Delta t} \boldsymbol{\epsilon}_\Theta^{t - \Delta t} (\x_{t - \Delta t}),
    \end{align*}
    we linearize $\boldsymbol{\epsilon}_\Theta^{t - \Delta t}(\x_{t - \Delta t})$ around $\boldsymbol{m}_{t - \Delta t} := \mathbb{E}[\x_{t - \Delta t}]$:
    \begin{align*}
        \boldsymbol{\epsilon}_\Theta^{t - \Delta t}(\x_{t - \Delta t}) 
        \approx \boldsymbol{E}_{t - \Delta t} 
        + \boldsymbol{J}_{t - \Delta t}(\x_{t - \Delta t} - \boldsymbol{m}_{t - \Delta t}) 
        + \boldsymbol{r}_t,
    \end{align*}
    where $\boldsymbol{E}_{t - \Delta t} = \boldsymbol{\epsilon}_\Theta^{t - \Delta t}(\boldsymbol{m}_{t - \Delta t})$, $\boldsymbol{J}_{t - \Delta t}$ is the Jacobian of $\boldsymbol{\epsilon}_\Theta^{t - \Delta t}$ evaluated at $\boldsymbol{m}_{t - \Delta t}$, and $\boldsymbol{r}_t$ is the Taylor remainder, independent of $\x_{t - \Delta t}$. Plugging this into the DDIM step yields:
    \begin{align*}
        \x_t &\approx 
        \left(\sqrt{\frac{\alpha_t}{\alpha_{t - \Delta t}}}\boldsymbol{I} + c_{t - \Delta t}\boldsymbol{J}_{t - \Delta t}\right)\x_{t - \Delta t}
        + c_{t - \Delta t}\boldsymbol{r}_t 
        + c_{t - \Delta t}\boldsymbol{E}_{t - \Delta t}
        - c_{t - \Delta t}\boldsymbol{J}_{t - \Delta t}\boldsymbol{m}_{t - \Delta t}.
    \end{align*}
    Since the last two terms are constant with respect to $\x_{t - \Delta t}$, their variance is zero. 
    Denoting $\boldsymbol{A}_t := \sqrt{\frac{\alpha_t}{\alpha_{t - \Delta t}}}\boldsymbol{I} + c_{t - \Delta t}\boldsymbol{J}_{t - \Delta t}$, taking the variance on both sides gives:
    \begin{align*}
        Var(\x_t) 
        &\approx Var\big( \boldsymbol{A}_t \x_{t - \Delta t} + c_{t - \Delta t}\boldsymbol{r}_t \big) \\
        &= \boldsymbol{A}_t Var(\x_{t - \Delta t}) \boldsymbol{A}_t^T + c_{t - \Delta t}^2 Var(\boldsymbol{r}_t),
    \end{align*}
    which is exactly \eqref{eq:recursive_formula_ndim}.
\end{proof}

To obtain a scalar measure of the overall uncertainty, we now consider the total variance $\TotVar(\x_t) := \Tr(\boldsymbol{\Sigma}_t)$. 
Using $\Tr(\boldsymbol{A}\boldsymbol{B}) = \Tr(\boldsymbol{B}\boldsymbol{A})$ and $\Tr(\boldsymbol{A}^T \boldsymbol{A}) = \|\boldsymbol{A}\|_F^2$, we can derive the following corollary.

\begin{corollary}\label{cor:totVar_recursion_ndim}
    The total variance $\v_t := \TotVar(\x_t)$ approximately satisfies the scalar affine recursion:
    \begin{align}\label{eq:totVar_recursion_ndim}
        \v_t \approx F_t\,\v_{t - \Delta t} + G_t,
    \end{align}
    where:
    \begin{align*}
        &F_t := \frac{1}{n}\Tr(\boldsymbol{A}_t^T\boldsymbol{A}_t)
        = \frac{1}{n}\|\boldsymbol{A}_t\|_F^2 > 0, \\
        &G_t := c_{t - \Delta t}^2\,\TotVar(\boldsymbol{r}_t) \ge 0.
    \end{align*}
\end{corollary}

Equation~\eqref{eq:totVar_recursion_ndim} generalizes the scalar recurrence \eqref{eq:recursive_formula} to arbitrary dimension $n$, preserving the same structure. 
By iteratively applying this recursion, we obtain a closed-form expression for $\v_T$, analogous to Theorem~\ref{teo:closed_formula} in the main text, and a continuous-time limit with exponential convergence properties.

\begin{theorem}\label{thm:ndim_closed_formula}
    Let $\v_t$ satisfy \eqref{eq:totVar_recursion_ndim} and assume that $\alpha(t)$, $\boldsymbol{J}_t$, and $\boldsymbol{r}_t$ are continuously differentiable in $t$. 
    Then, as $\Delta t \to 0$, it holds:
    \begin{align}\label{eq:ndim_limit_variance}
        \lim_{\Delta t \to 0} \v_T 
        = e^{-\int_0^T \lambda(s)ds} 
        + \int_0^T e^{-\int_0^s \lambda(\kappa)d\kappa}\,\gamma(s)\,ds,
    \end{align}
    where:
    \begin{align*}
        &\lambda(t) 
        = - \partial_t \log \alpha(t) 
        + \frac{1}{n}\Tr(\boldsymbol{J}_t)\,\alpha'(t)
        \left(\frac{\sqrt{1 - \alpha(t)}}{\alpha(t)} 
        + \frac{1}{\sqrt{1 - \alpha(t)}}\right), \\
        &\gamma(t) 
        = \left(
        \frac{\alpha'(t)\sqrt{1 - \alpha(t)}}{2\alpha(t)}
        + \frac{\alpha'(t)}{2\sqrt{1 - \alpha(t)}}
        \right)
        \TotVar(\boldsymbol{r}_t).
    \end{align*}
\end{theorem}

\begin{proof}
    The proof follows exactly the same steps as Theorem~\ref{teo:closed_formula} for the scalar case. 
    Taylor expanding $\sqrt{\alpha(t)}$ and $\sqrt{1-\alpha(t)}$ around $t-\Delta t$, and using:
    \begin{align*}
        \sqrt{\frac{\alpha(t)}{\alpha(t-\Delta t)}} 
        &= 1 - \frac{\alpha'(t - \Delta t)}{2\alpha(t-\Delta t)} \Delta t + O(\Delta t^2), \\
        c_{t - \Delta t} 
        &= \left( \frac{1}{2\sqrt{1 - \alpha(t - \Delta t)}} + \frac{\sqrt{1 - \alpha(t - \Delta t)}}{2\alpha(t - \Delta t)}\right) \alpha'(t - \Delta t) \Delta t + O(\Delta t^2),
    \end{align*}
    yields:
    \begin{align*}
        \boldsymbol{A}_t 
        &= \boldsymbol{I} 
        + \Delta t\left[-\frac{1}{2}\partial_t\!\log\alpha(t-\Delta t)\,\boldsymbol{I}
        + q_{t-\Delta t}\,\boldsymbol{J}_{t-\Delta t}\right] + O(\Delta t^2),
    \end{align*}
    where 
    \(
    q_{t-\Delta t} = \frac{\alpha'(t - \Delta t)}{2\alpha(t - \Delta t)}\sqrt{1-\alpha(t-\Delta t)} 
    + \frac{\alpha'(t - \Delta t)}{2\sqrt{1-\alpha(t-\Delta t)}}.
    \)
    Therefore:
    \begin{align*}
        F_t 
        &= \frac{1}{n}\Tr(\boldsymbol{A}_t^T\boldsymbol{A}_t)
        = 1 + \lambda(t-\Delta t)\Delta t + O(\Delta t^2), \\
        G_t 
        &= c_{t - \Delta t}^2\,\TotVar(\boldsymbol{r}_t)
        = \gamma(t-\Delta t)\Delta t + O(\Delta t^2),
    \end{align*}
    with $\lambda$ and $\gamma$ defined as above. 
    Unrolling the recursion \eqref{eq:totVar_recursion_ndim} leads to a product–sum expression identical to \eqref{eq:closed_formula}, and standard Euler-sum arguments show that the discrete products $\prod F_t$ converge monotonically from below to $e^{\int\lambda}$, while the additive term converges to the integral in \eqref{eq:ndim_limit_variance}. 
    Monotonicity of $\{\v_{N\Delta t}\}_{N\in\mathbb{N}}$ follows from the same argument used in the scalar case: since $F_t>0$ and $G_t\ge 0$, each refinement step increases $\v_t$, completing the proof.
\end{proof}

\end{document}